\documentclass[sigconf]{acmart}

\def\BibTeX{{\rm B\kern-.05em{\sc i\kern-.025em b}\kern-.08emT\kern-.1667em\lower.7ex\hbox{E}\kern-.125emX}}
\usepackage{amsmath}
\usepackage{bm}
\usepackage{float}
\usepackage{graphicx}
\usepackage{multirow}
\usepackage{tikz}
\usetikzlibrary{bayesnet}

\DeclareMathOperator*{\argmax}{arg\,max}

\copyrightyear{2020}
\acmYear{2020}
\setcopyright{iw3c2w3}
\acmConference[WWW '20]{Proceedings of The Web Conference 2020}{April 20--24, 2020}{Taipei, Taiwan}
\acmBooktitle{Proceedings of The Web Conference 2020 (WWW '20), April 20--24, 2020, Taipei, Taiwan}
\acmPrice{}
\acmDOI{10.1145/3366423.3380123}
\acmISBN{978-1-4503-7023-3/20/04}

\begin{document}

%
\title{Open Knowledge Enrichment for Long-tail Entities}

%
\author{Ermei Cao}
\authornote{Equal contributors}
\affiliation[obeypunctuation=true]{
	\department{State Key Laboratory for Novel Software Technology}\\
	\institution{Nanjing University},
	\country{China}}
\email{emcao.nju@gmail.com}

\author{Difeng Wang}
\authornotemark[1]
\affiliation[obeypunctuation=true]{
	\department{State Key Laboratory for Novel Software Technology}\\
	\institution{Nanjing University},
	\country{China}}
\email{dfwang.nju@gmail.com}

\author{Jiacheng Huang}
\affiliation[obeypunctuation=true]{
	\department{State Key Laboratory for Novel Software Technology}\\
	\institution{Nanjing University},
	\country{China}}
\email{jchuang.nju@gmail.com}

\author{Wei Hu}
\authornote{Corresponding author}
\affiliation[obeypunctuation=true]{
	\department{State Key Laboratory for Novel Software Technology}\\
	\department{National Institute of Healthcare Data Science}\\
	\institution{Nanjing University},
	\country{China}}
\email{whu@nju.edu.cn}

%
\begin{abstract}
Knowledge bases (KBs) have gradually become a valuable asset for many AI applications. While many current KBs are quite large, they are widely acknowledged as incomplete, especially lacking facts of long-tail entities, e.g., less famous persons. Existing approaches enrich KBs mainly on completing missing links or filling missing values. However, they only tackle a part of the enrichment problem and lack specific considerations regarding long-tail entities. In this paper, we propose a full-fledged approach to knowledge enrichment, which predicts missing properties and infers true facts of long-tail entities from the open Web. Prior knowledge from popular entities is leveraged to improve every enrichment step. Our experiments on the synthetic and real-world datasets and comparison with related work demonstrate the feasibility and superiority of the approach.
\end{abstract}



%

%
\maketitle

\section{Introduction}
\label{sect:intro}

\begin{figure}
\centering
\includegraphics[width=.75\columnwidth]{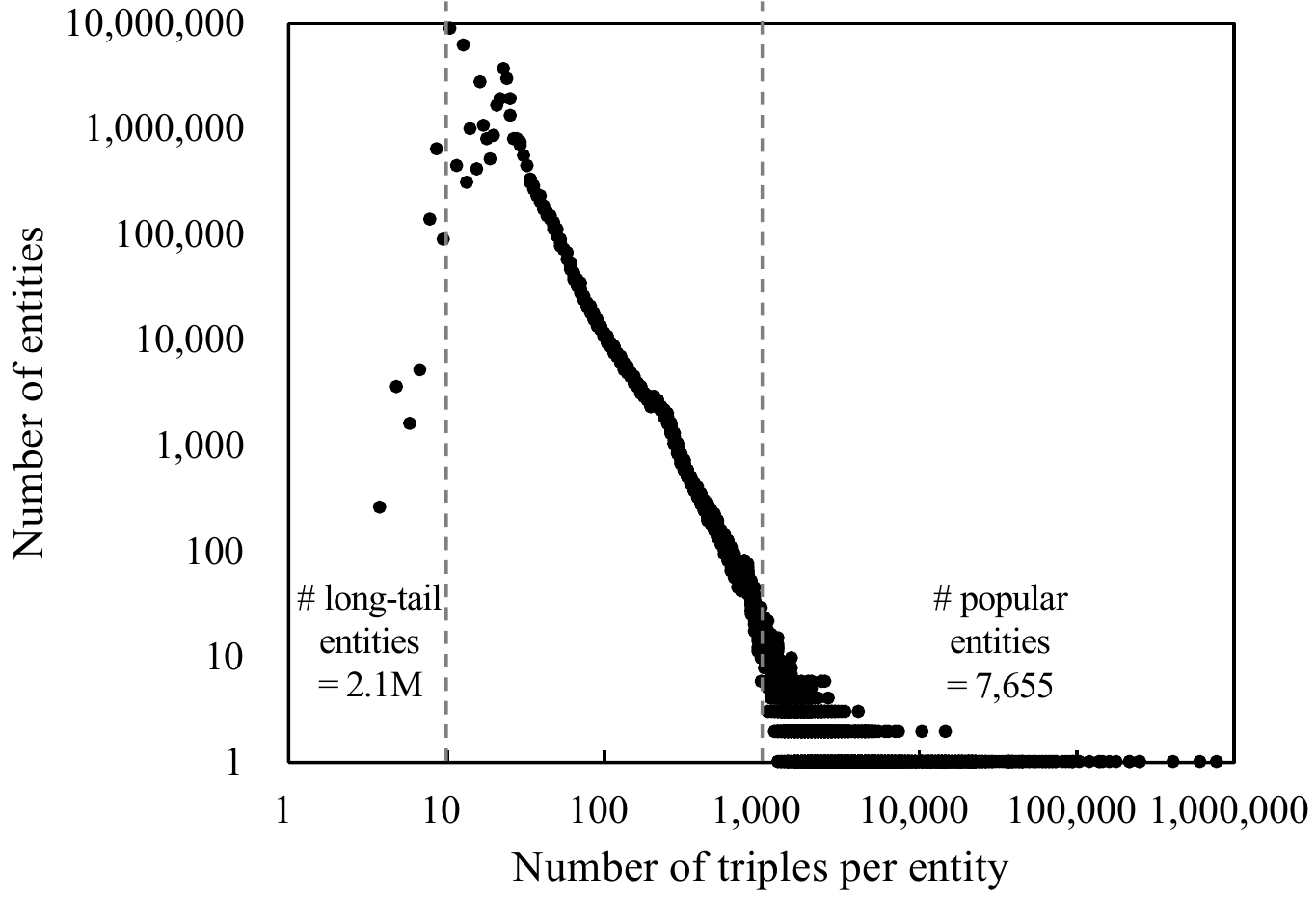}
\Description{Power-law distribution}
\caption{Distribution of number of entities versus number of triples per entity (in log-log scale)}
\label{fig:freebase}
\end{figure}

The last few years have witnessed that knowledge bases (KBs)~have become a valuable asset for many AI applications, such as semantic search, question answering and recommender systems. Some existing KBs, e.g., Freebase, DBpedia, Wikidata and Probase, are very large, containing millions of entities and billions of facts, where a fact is organized as a triple in the form of $\langle entity, property, value \rangle$. However, it has been aware that the existing KBs are likely to have high recall on the facts of popular entities (e.g., celebrities, famous places and award-winning works), but are overwhelmingly incomplete on less popular (e.g., \emph{long-tail}) entities \cite{KnowledgeVault,KnowMore,KBRecall}. For instance, as shown in Figure~\ref{fig:freebase}, around 2.1 million entities in Freebase have less than 10 facts per entity, while 7,655 entities have more than one thousand facts, following the so-called \emph{power-law} distribution. 

Among those long-tail entities, some just lack facts in KBs rather than in the real world. The causes of the incompleteness are manifold. First, the construction of large KBs typically relies on soliciting contributions from human volunteers or distilling knowledge from ``cherry-picked'' sources like Wikipedia, which may yield a limited coverage on frequently-mentioned facts \cite{KnowledgeVault}. Second, some formerly unimportant or unknown entities may rise to fame suddenly, due to the dynamics of this ever-changing world \cite{EmergEntity}. However, current KBs may not be updated in time. Because the Web has become the main source for people to access information nowadays, the goal of this paper is to conjecture what facts about long-tail entities are missing, as well as extract and infer true facts from various Web sources. We believe that enriching long-tail entities with uncovered facts from the open Web is vital for building more complete KBs. 

\textbf{State-of-the-art and limitations.} As investigated in \cite{HowMuchTriple}, the cost of curating a fact manually is much more expensive than that of automatic creation, by a factor of 15 to 250. Due to the vast scale of long-tail entities in KBs and accessible knowledge on the Web, automation is inevitable. Existing approaches address this problem from various angles \cite{KBPSurvey,KGRefineSurvey,KGEmbedSurvey}, however, we argue that they may have the following two limitations:

First, existing approaches only deal with a part of the knowledge enrichment problem, such as recommending properties to entities \cite{PredSug,Obligatory}, predicting missing links between entities \cite{TransE,ConvE,ConMask,RotatE} and verifying the truths of facts \cite{Facty}. Also, the KB population (KBP) or slot filling approaches usually assume that the target properties are given beforehand and extract values from free text \cite{JointKBP,SFOverview} or structured Web tables \cite{Profile,NovelTable,KnowMore}. To the best of our knowledge, none of them can accomplish open knowledge enrichment alone.

Second, most approaches lack considerations for long-tail entities. Due to the lack of facts about long-tail entities, the link prediction approaches may not learn good embeddings for them. Similarly, the KBP approaches would be error-sensitive for incidentally-appeared entities, as they cannot handle errors or exceptions well. We note that a few works have begun to study the long-tail phenomenon in KBs, but they tackle different problems, e.g., linking long-tail entities to KBs \cite{LongTNews}, extracting long-tail relations \cite{LongTRelExt} and verifying facts for long-tail domains \cite{Facty}.

\textbf{Our approach and contributions.} To address the above limitations, we propose \emph{OKELE}, a full-fledged approach to enrich long-tail entities from the open Web. OKELE works based on the idea that we can infer the missing knowledge of long-tail entities by comparing with similar popular entities. For instance, to find out what a person lacks, we can see what other persons have. We argue that, this may not be the best solution, but it is sufficiently intuitive and very effective in practice. 

Specifically, given a long-tail entity, OKELE aims to search the Web to find a set of true facts about it. To achieve this, we deal with several challenges: First, the candidate properties for a long-tail entity can be vast. We construct an entity-property graph and propose a graph neural network (GNN) based model to predict appropriate properties for it. Second, the values of a long-tail entity are scattered all over the Web. We consider various types of Web sources and design corresponding extraction methods to retrieve them. Third, the extracted facts from different sources may have conflicts. We propose a probabilistic graphical model to infer the true facts, which particularly considers the imbalance between small and large sources. Note that, during the whole process, OKELE makes full use of popular entities to improve the enrichment accuracy.

The main contributions of this paper are summarized as follows:
\begin{itemize}
\item We propose a full-fledged approach for open knowledge enrichment on long-tail entities. As far as we know, this is the first work attempting to solve this problem. 

\item We propose a novel property prediction model based on GNNs and graph attention mechanism, which can accurately predict the missing properties of long-tail entities by comparison of similar popular entities. 

\item We explore various semi-structured, unstructured and structured Web sources. For each type of sources, we develop the corresponding extraction method, and use popular entities to find appropriate sources and refine extraction methods. 

\item We present a new fact verification model based on a probabilistic graphical model with conjugate priors, which infers the true facts of long-tail entities by incorporating confidence interval estimators of source reliability and prior knowledge from popular entities. 

\item We conduct both synthetic and real-world experiments. Our results demonstrate the effectiveness of OKELE, and also show that the property prediction and fact verification models significantly outperform competitors. 
\end{itemize}


\section{Overview of the Approach}
\label{sect:overview}

In this paper, we deal with RDF KBs such as Freebase, DBpedia and Wikidata. A KB is defined as a 5-tuple $KB=(E,P,T,L,F)$, where $E,P,T,L,F$ denote the sets of entities, properties, classes, literals and facts, respectively. A fact is a triple of the form $f=\langle e,p,x \rangle \in E\times P\times (E\cup T\cup L)$, e.g., $\langle \textit{Erika\_Harlacher}, \textit{profession}, \textit{Voice\_Actor} \rangle$. Moreover, properties can be divided into relations and attributes. Usually, it is hard to strictly distinguish \emph{long-tail} entities and \emph{non-long-tail} entities. Here, we roughly say that an entity is a long-tail entity if it uses a small number of distinct properties (e.g., $\le 5$).

Given a long-tail entity $e$ in a KB, \emph{open knowledge enrichment} aims at tapping into the masses of data over the Web to infer the missing knowledge (e.g., properties and facts) of $e$ and add it back to the KB. A Web source, denoted by $s$, makes a set of claims, each of which is in the form of $c=(f,s,o)$, where $f,s,o$ are a fact, a source and an observation, respectively. In practice, $o$ often takes a confidence value about how confident $f$ is present in $s$ according to some extraction method. Also, $f$ has a label $l_f \in \{\text{True},\text{False}\}$. 

Figure~\ref{fig:overview} shows the workflow of our approach, which accepts a long-tail entity as input and conducts the following three steps:

\begin{figure}
\centering
\includegraphics[width=.72\columnwidth]{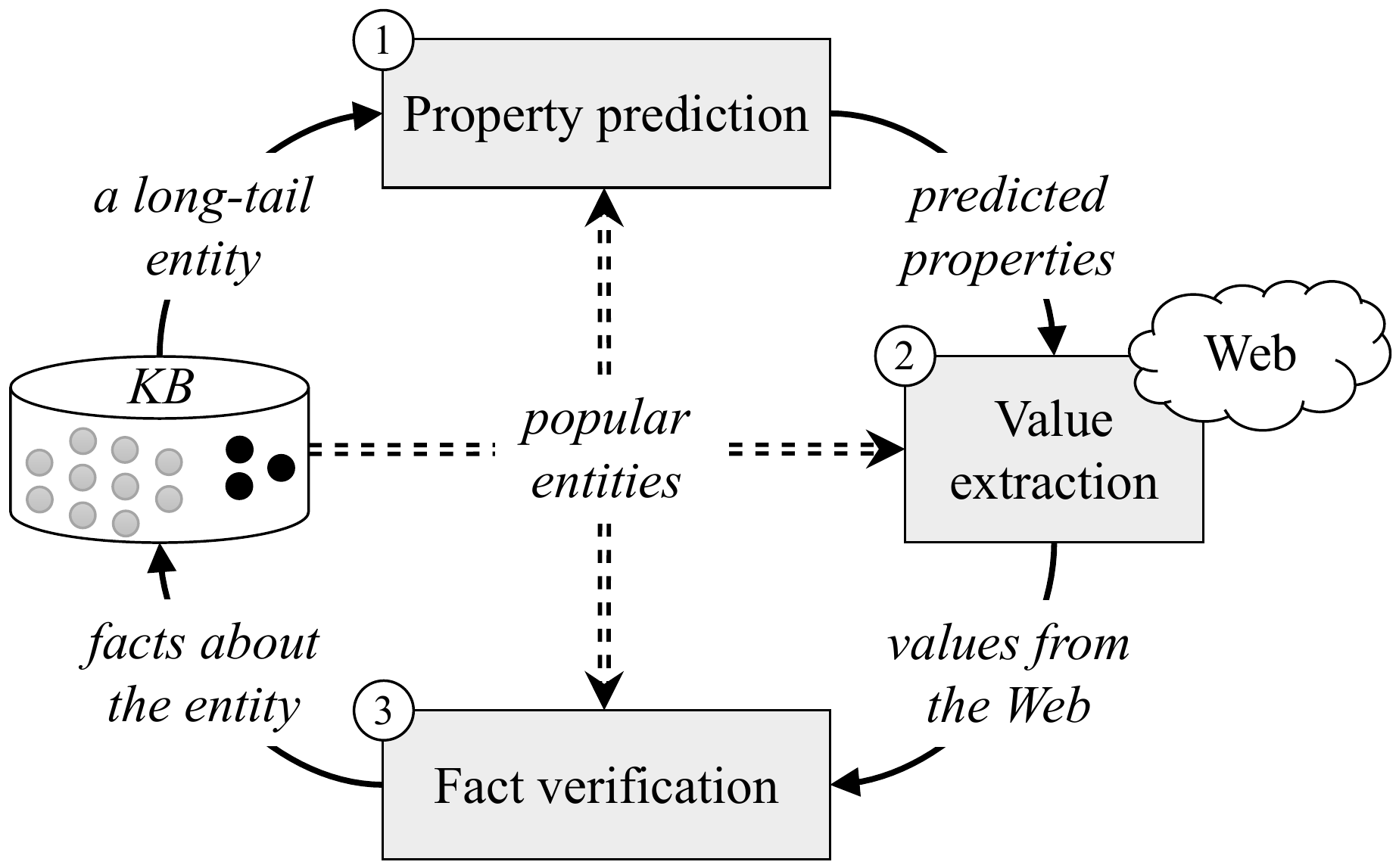}
\Description{Workflow}
\caption{Overview of the approach}
\label{fig:overview}
\end{figure}

\begin{enumerate}
\item \textbf{Property prediction.} Based on the observations that similar entities are likely to share overlapped properties and popular entities have more properties, we resort to \emph{similar popular entities}. Our approach first creates an entity-property graph to model the interactions between entities and properties. Then, it employs an attention-based GNN model to 
predict the properties of the long-tail entity.

\item \textbf{Value extraction.} For each predicted property of the long-tail entity, our approach extracts the corresponding values from the Web. To expand the coverage and make full use of the redundancy of Web data, we leverage various types of Web sources, including semi-structured vertical websites, unstructured plain text in Web content and structured HTML tables. Prior knowledge from popular entities is used to improve the template-based extraction from vertical websites and the distantly-supervised extraction from text.

\item \textbf{Fact verification.} Our approach employs an efficient probabilistic graphical model to estimate the probability of each fact being true, based on the observations from various Web sources. To tackle the skewed number of claims caused by the long-tail phenomenon, our approach adopts an effective estimator to deal with the effect of source claim size. Moreover, prior knowledge from popular entities is leveraged to guide the estimation of source reliability. Finally, the verified facts are added into the KB to enrich the long-tail entity.
\end{enumerate}

\section{Property Prediction}
\label{sect:property}

To leverage similar popular entities to infer the missing properties of long-tail entities, the main difficulties lie in how to model the interactions between entities and properties and make an accurate prediction from a large number of candidates. In this section, we present a new property prediction model based on GNNs~\cite{GNN,GCN} and graph attention mechanism~\cite{GAT}. 

\subsection{Entity-Property Graph Construction}

We build \emph{entity-property graphs} to model the interactions between entities and properties and the interactions between similar entities. An entity-property graph consists of two types of nodes, namely \emph{entity nodes} and \emph{property nodes}, and two types of edges: (i) \emph{entity-property edges} are used to model the interactions between entities and properties. We create an edge between an entity node and a property node if the entity uses the property; and (ii) \emph{entity-entity edges} are used to model the interactions between similar entities. We create an edge between an entity node and each of its top-$k$ similar entity nodes.

Given a $KB=$ $(E,P,T,L,F)$, for any two entities $e_1,e_2\in E$, we consider three aspects of similarities between them. Note that, in practice, we sample a small subset of $KB$ for model training.

\begin{itemize}

\item \textbf{Type-based similarity} considers the number of types (i.e., classes) that $e_1,e_2$ have in common. To emphasize that some types are more informative than others (e.g., \textit{actor} versus \textit{person}), we further employ a weighted scoring function:
\begin{align}
\text{sim}^t_{1,2} = \frac{2 \sum_{t\in T_{e_1} \cap\,T_{e_2}} q_t}{\sum_{t_1\in T_{e_1}} q_{t_1} + \sum_{t_2\in T_{e_2}} q_{t_2}},
\end{align}
where $q_t=\frac{|E|}{|\{e\in E \,:\, e \text{ has type } t\}|}$.  $T_e$ denotes the set of types that $e$ directly defines (i.e., without subclass reasoning). 

\item \textbf{Property-based similarity} measures the similarity based on the number of common properties used by $e_1,e_2$:
\begin{align}
\resizebox{.9\columnwidth}{!}{$\begin{aligned}
\text{sim}^p_{1,2} = \frac{2\,|\{p\in P : \langle e_1,p,x \rangle\in F\} \bigcap\, \{p\in P : \langle e_2,p,x' \rangle\in F\}|}{|\{p\in P : \langle e_1,p,x \rangle\in F\}| + |\{p\in P : \langle e_2,p,x' \rangle\in F\}|}.
\end{aligned}$}
\end{align}

\item \textbf{Value-based similarity} calculates the number of values that $e_1,e_2$ both have. Analogous to the type-based similarity, we emphasize more informative values:
\begin{align}
\text{sim}^v_{1,2} = \frac{2 \sum_{v\in V_{e_1} \cap\,V_{e_2}} \text{info}(v)}{\sum_{v_1\in V_{e_1}} \text{info}(v_1) + \sum_{v_2\in V_{e_2}} \text{info}(v_2)},
\end{align}
where $\text{info}(v)=\log\frac{|E|}{|\{e\in E \,:\, e \text{ has value } v\}|}$ \cite{FACES}. $V_e$ denotes the set of values that $e$ has. For entities and classes, we directly use the URLs, and for literals, we use the lexical forms.

\end{itemize}

The \emph{overall} similarity between $e_1$ and $e_2$ is obtained by linearly combining the above three similarities:
\begin{align}
\text{sim}_{1,2} = \alpha_1\,\text{sim}^t_{1,2}+ \alpha_2\,\text{sim}^p_{1,2}+ \alpha_3\,\text{sim}^v_{1,2},
\end{align} 
where $\alpha_1,\alpha_2,\alpha_3\in [0,1]$ are weighting factors s.t. $\alpha_1+\alpha_2+\alpha_3=1$.

\subsection{Attention-based GNN Model}

\begin{figure*}
\centering
\includegraphics[width=.75\textwidth]{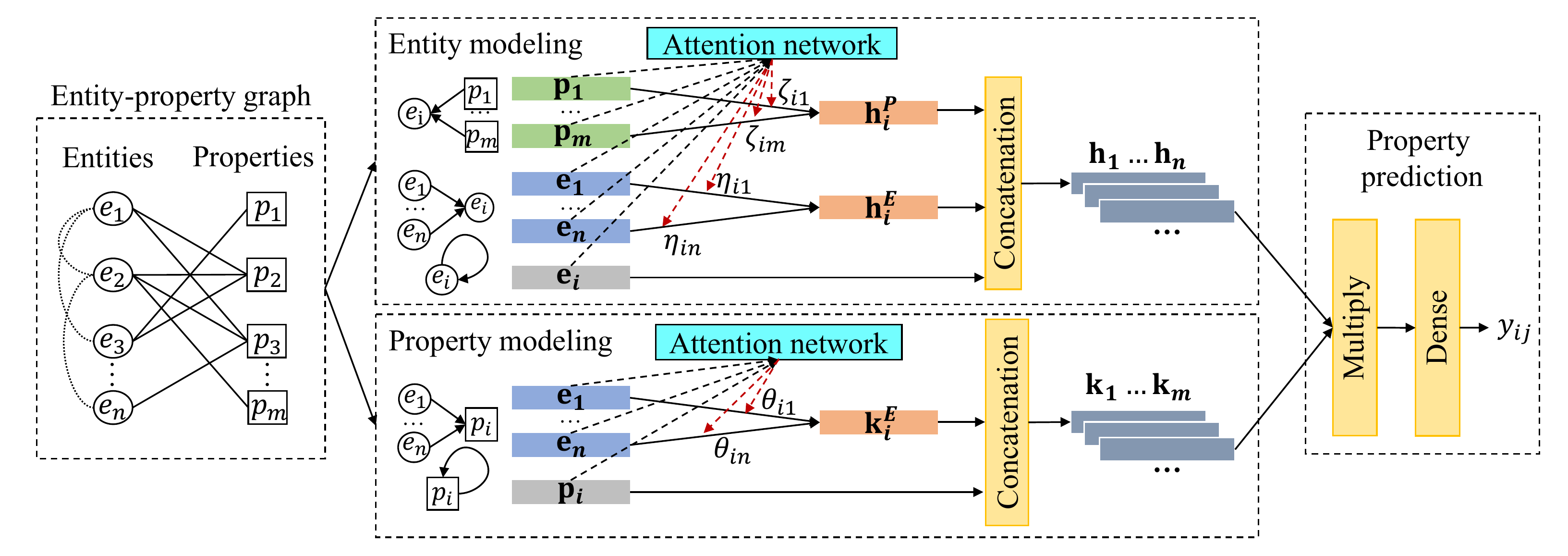}
\Description{Property prediction model}
\caption{Attention-based GNN model for property prediction}
\label{fig:pp_gnn}
\end{figure*}

Figure~\ref{fig:pp_gnn} shows our attention-based GNN model for property prediction. Below, we present its main modules in detail. For notations, we use boldface lowercase and uppercase letters to denote vectors and matrices, respectively.

\textbf{Entity modeling.} For each entity $e_i$, entity modeling learns a corresponding latent vector representation $\mathbf{h}_i$, by aggregating three kinds of interactions, namely entity-property interactions (denoted by $\mathbf{h}_i^P$), entity-entity interactions ($\mathbf{h}_i^E$) and itself:
\begin{align}
\mathbf{h}_i=\mathbf{h}_i^P+\mathbf{h}_i^E+\sigma(\mathbf{W}\,\mathbf{e}_i+\mathbf{b}),
\end{align}
where $\mathbf{e}_i$ denotes the embedding of $e_i$. $\sigma()$ is the activation function. $\mathbf{W}$ and $\mathbf{b}$ are the weight matrix and bias vector, respectively. 

Specifically, entity-property interactions aggregate information from its neighboring property nodes:
\begin{align}
	\mathbf{h}_i^P = \sigma\Big( \mathbf{W} \big( \sum_{j \in N_i^P}\zeta_{ij}\,\mathbf{p}_j \big) + \mathbf{b} \Big),
\end{align}
where $\mathbf{p}_j$ denotes the embedding of property $p_j$. $\zeta_{ij}$ is the interaction coefficient between $e_i$ and $p_j$, and $N_i^P$ is the neighboring property node set of $e_i$.  

Entity-entity interactions aggregate information from its neighboring entity nodes:
\begin{align}
\mathbf{h}_i^E = \sigma\Big( \mathbf{W} \big( \sum_{j \in N_i^E}\eta_{ij}\,\mathbf{e}_j \big) + \mathbf{b} \Big), 
\end{align}
where $\eta_{ij}$ is the interaction coefficient between $e_i $ and $e_j$, and $N_i^E$ is the neighboring entity node set of $e_i$.

There are two ways of calculating the interaction coefficients $\zeta_{ij}$ and $\eta_{ij}$. One fixes $\zeta_{ij}=\frac{1}{|N_i^P|},\eta_{ij}=\frac{1}{|N_i^E|}$ under the assumption that all property/entity nodes contribute equally. However, this may not be optimal, because (i) different properties have different importance to one entity. For example, \textit{birth\_date} is less specific than \textit{written\_book} for an author; and (ii) for an entity, its similar neighboring entities can better model that entity. Thus, the other way lets the interactions have uneven contributions to the latent vector representations of entities. Here, we employ the graph attention mechanism \cite{GAT} to make the entity pay more attention to the interactions with more relevant properties/entities, which are formulated as follows:
\begin{align}
	\zeta_{ij} &= \mathbf{w}_2^T\,\sigma(\mathbf{W}_1\,[\mathbf{p}_j;\mathbf{e}_i] + \mathbf{b}_1) + b_2, \\
	\eta_{ij} &= \mathbf{w}_2^T\,\sigma(\mathbf{W}_1\,[\mathbf{e}_j;\mathbf{e}_i] + \mathbf b_1) + b_2,
\end{align}
where $[; ]$ is the concatenation operation. We normalize the coefficients using the softmax function, which can be interpreted as the importance of $p_j$ and $e_j$ to the latent vector representation of $e_i$.

\textbf{Property modeling.} For each property $p_i$, property modeling learns a corresponding latent vector representation $\mathbf{k}_i$, by aggregating two kinds of interactions, namely property-entity interactions ($\mathbf{k}_i^E$) and itself:
\begin{align}
\mathbf{k}_i=\mathbf{k}_i^E+\sigma(\mathbf{W}\,\mathbf{p}_i+\mathbf{b}).
\end{align}

Specifically, property-entity interactions aggregate information from its neighboring entity nodes:
\begin{align}
\mathbf{k}_i^E = \sigma\Big( \mathbf{W} \big( \sum_{j \in M_i^E}\theta_{ij}\,\mathbf{e}_j \big) + \mathbf{b} \Big),
\end{align}
where $\theta_{ij}$ is the interaction coefficient between $p_i$ and $e_j$, and $M_i^E$ is the neighboring entity node set of $p_i$.  

Similarly, we use the graph attention mechanism to refine $\theta_{ij}$:
\begin{align}
	\theta_{ij} &= \mathbf{w}_2^T\,\sigma(\mathbf{W}_1\,[\mathbf{e}_j;\mathbf{p}_i] + \mathbf{b}_1) + b_2, 
\end{align}
which is normalized by softmax as well.

\textbf{Property prediction.} For entity $e_i$ and property $p_j$, we first multiply them as $\mathbf{g}_0 = [\mathbf{h}_i \cdot \mathbf{k}_j]$, which is then fed into a multi-layer perceptron (MLP) to infer the probability $y_{ij}$ that $e_i$ has $p_j$ using the sigmoid function:
\begin{align}
\begin{aligned}
	\mathbf{g}_1 = \sigma(\mathbf{W}_0\,\mathbf{g}_0 &+ \mathbf{b}_0), \cdots, \mathbf{g}_l = \sigma(\mathbf{W}_{l-1}\,\mathbf{g}_{l-1} + \mathbf{b}_{l-1}), \\
	y_{ij} &= \text{sigmoid}(\mathbf{w}_l^T\,\mathbf{g}_l + b_l),
\end{aligned}
\end{align}
where $l$ is the number of hidden layers.

\textbf{Model training.} We define the loss function as follows:
\begin{align}
\text{loss} = -\sum_{i=1}^{n}\sum_{j=1}^{m}\Big[y_{ij}^*\,\log(y_{ij}) + (1-y_{ij}^*)\,\log(1-y_{ij})\Big], 
\end{align}
where $y_{ij}^*$ is the true label of $e_i$ having $p_j$, and $n,m$ are the numbers of entities and properties in the training set. 
To optimize this loss function, we use Adam optimizer \cite{Adam} and the polynomial decay strategy to adjust the learning rate. The model parameter complexity is $\mathrm{O}\big((n + m) d_1 + l d_1^2 + d_1 d_2\big)$, where $d_1$ is the dimension of embeddings and latent vectors, and $d_2$ is the size of attentions.

\begin{figure}[!t]
\centering
\includegraphics[width=\columnwidth]{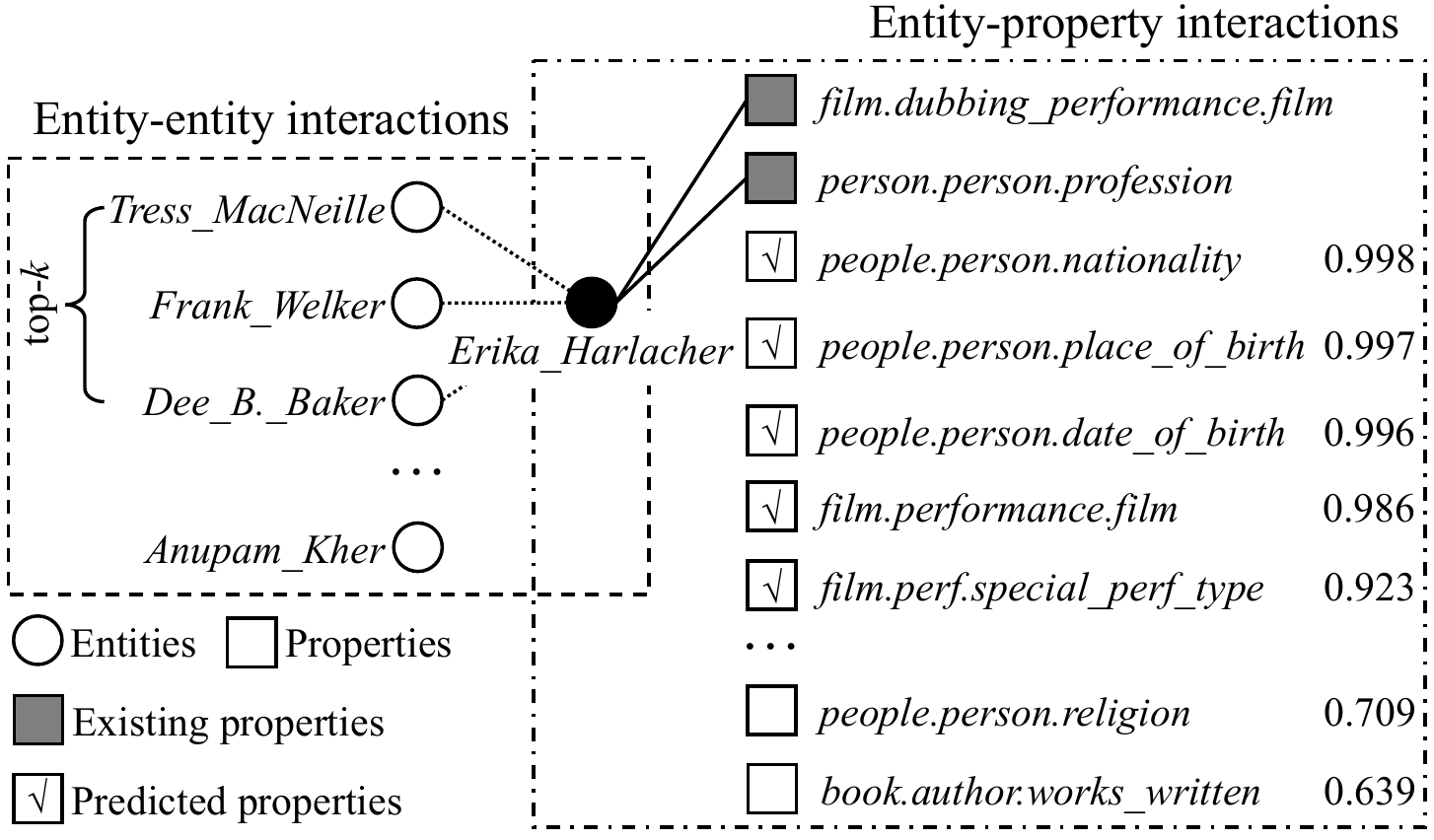}
\Description{Example 1}
\caption{Example of property prediction}
\label{fig:example1}
\end{figure}

\begin{example}
Let us see the example depicted in Figure~\ref{fig:example1}. \textit{Erika\_} \textit{Halacher} is an American voice actress. She is a long-tail entity in Freebase with two properties \textit{film.dubbing\_performance.film} and \textit{person.person.profession}. Also, according to the similarity measures, \textit{Tress\_MacNeille}, \textit{Frank\_Welker} and \textit{Dee\_B.\_Baker} are ranked as its top-3 similar entities. The attention-based GNN model predicts a group of properties for \textit{Erika\_Halacher}, some of which are general, e.g., \textit{people.person.nationality}, while others are customized, such as \textit{film.performance.special\_performance\_type}. $\Box$
\end{example}

\section{Value Extraction}
\label{sect:value}

Given an entity and a set of predicted properties, we aim to search the corresponding value collections on the Web, where each value collection is associated with an entity-property pair. However, extracting values for long-tail entities from the Web is hard. On one hand, there are many different types of long-tail entities and their properties are pretty diverse and sparse. On the other hand, a large portion of Web data is un/semi-structured and scattered, there is no guarantee for their veracity. Thus, to improve the coverage of value collections and make full use of the redundancy of Web data, we consider \emph{semi-structured vertical websites}, \emph{unstructured plain text} and \emph{structured data}.  

\subsection{Extraction from Vertical Websites}

Vertical websites contain high-quality knowledge for entities of specific types, e.g., IMDB\footnote{\url{https://www.imdb.com}} is about actors and movies. As found in~\cite{CERES}, a vertical website typically consists of a set of entity detail pages generated from a template or a set of templates. Each detail page describes an entity and can be regarded as a DOM tree. Each node in the DOM tree can be reached by an absolute XPath. The detail pages using the same template often share a common structure and placement of content. 

\textbf{Method.} We propose a two-stage method to extract values from vertical websites. First, we leverage \emph{popular entities} in the previous step to find appropriate vertical websites. For a popular entity, we use its name and type(s) as the query keywords to vote and sort vertical websites through a search engine. 

Then, we use the \emph{known facts} of popular entities to learn one or more XPaths for each property. In many cases, the XPaths for the same property are likely to be similar. For example, for property \textit{date\_of\_birth}, the XPaths in \textit{Tom\_Cruise} and \textit{Matt\_Damon} detail pages of IMDB are the same. For the case that a page puts multiple values of the same property together, e.g., in the form of a list or table, we merge the multiple XPaths to a generic one. Also, we use CSS selectors to enhance the extraction by means of HTML tags such as \textit{id} and \textit{class}. For example, the CSS selector for \textit{date\_of\_birth} is always ``\#name-born-info > time" in IMDB. Thus, for each vertical website, we can learn a set of XPath-property mappings, which are used as templates to extract values for long-tail entities.

\textbf{Implementation.} We employ Google as the search engine, and sample 200 popular entities for each type of entities. According to our experiments, the often used vertical websites include IMDB, Discogs,\footnote{\url{https://www.discogs.com}, $^3$\url{http://www.goodreads.com}, $^4$\url{https://www.drugbank.ca}, $^5$\url{http://peakbagger.com}} GoodReads,\footnotemark[3] DrugBank,\footnotemark[4] and Peakbagger.\footnotemark[5]

\subsection{Extraction from Plain Text}

We use \emph{closed IE} (information extraction) to extract knowledge from unstructured text. For open IE~\cite{OpenIE}, although it can extract facts on any domain without a vocabulary, the extracted textual facts are hard to be matched with KBs. We do not consider it currently. Again, we leverage popular entities to improve the extraction accuracy. 

\textbf{Method.} We first perform natural language processing (NLP) jobs, e.g., named entity recognition (NER) and coreference resolution, on each document. We conduct entity linking using the methods described in \cite{EntityLink}. Then, we leverage \emph{distant supervision} \cite{DS} for training. Distantly-supervised relation extraction assumes that, if two entities have a relation in a KB, then all sentences that contain these two entities hold this relation. However, this would generate some wrongly-labeled training data. To further reduce the influence of errors and noises, we use a relation extraction model with multi-instance learning \cite{OpenNRE}, which can dynamically reduce the weights of those noisy instances.

\textbf{Implementation.} To get the text for each entity, we use the snippets of its top-10 Google search results and its Wikipedia page. We leverage Stanford CoreNLP toolkit~\cite{StanfordNLP} together with DBpedia-Spotlight~\cite{Spotlight} and n-gram index search for NER. Other NLP jobs are done with Stanford CoreNLP toolkit. For relation extraction, we use the Wikipedia pages of popular entities as the annotated corpus for distant supervision and implement the sentence-level attention over multiple instances with OpenNRE~\cite{OpenNRE}.

\subsection{Extraction from Structured Data}

For structured data on the Web, we mainly consider relational Web tables and Web markup data. Previous studies \cite{Profile,NovelTable} have shown that Web tables contain a vast amount of structured knowledge about entities. Additionally, there are many webpages where their creators have added structured markup data, using the \textit{\url{schema.org}} vocabulary along with the Microdata, RDFa, or JSON-LD formats.

\textbf{Method.} For relational Web tables, the extraction method consists of three phases: (i) \emph{table search}, which uses the name and type of a target entity to find related tables; (ii) \emph{table parsing}, which retrieves entity facts from the tables. Following \cite{Facty}, we distinguish vertical tables and horizontal tables. A vertical table, e.g., a Wikipedia infobox, usually describes a single entity by two columns, where the first column lists the properties while the second column provides the values. We can extract facts row by row. A horizontal table often contains several entities, where each row describes an entity, each column represents a property and each cell gives the corresponding value. We identify which row refers to the target entity using string matching, and extract the table header or the first non-numeric row as properties; and (iii) \emph{schema matching}, which matches table properties to the ones in a KB. We compare the labels of properties after normalization and also extend labels with synonyms in WordNet.

For Web markup data, as the properties from schema.org vocabulary are canonical, and the work in \cite{Voldermortkg} has shown that string comparison between labels of markup entities is an efficient way for linking coreferences, we reuse the entity linking and property matching methods as aforementioned.

\setcounter{footnote}{5}

\textbf{Implementation.} We collect the English version of WikiTables \cite{WikiTables} and build a full-text index for search based on Lucene. We also use the online interface of Google Web tables.\footnote{\url{https://research.google.com/tables}, $^7$\url{http://webdatacommons.org/structureddata}} For Web markup data, we retrieve from the Web Data Commons Microdata corpus.\footnotemark[7]

\section{Fact Verification}
\label{sect:fact}

Due to the nature of the Web and the imperfection of the extraction methods, conflicts often exist in the values collected from different sources. Among the conflicts, which one(s) represent the truth(s)? Facing the daunting data scale, expecting humans to check all facts is unrealistic, so our goal is to algorithmically verify their veracity.

As the simplest model, majority voting treats the facts claimed by the majority as correct, but it fails to distinguish the reliability of different sources, which may lead to poor performance when the number of low-quality sources is large. A better solution evaluates sources based on the intuition that high-quality sources are likely to provide more reliable facts. However, the reliability is usually prior unknown. Moreover, stepping into the era of Web 2.0, all end users can create Web content freely, which causes that many sources just provide several claims about one or two entities, while only a few sources make plenty of claims about many entities, i.e., the so-called \emph{long-tail} phenomenon~\cite{CATD}. It is very difficult to assess the reliability of those ``small" sources accurately, and an inaccurate estimate would impair the effectiveness of fact verification. To tackle these issues, we propose a novel probabilistic graphical model.

\subsection{Probabilistic Graphical Model}
\label{subsect:pgm} 

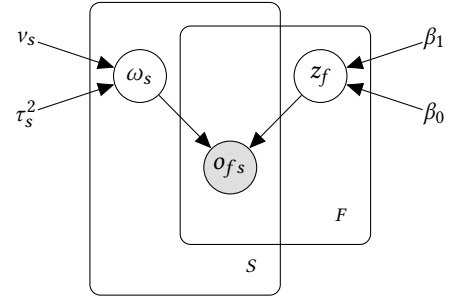
\begin{figure}
  \begin{tikzpicture}
    \node[obs] (ofs) {$o_{fs}$} ; %
    \node[latent, above=0.5cm of ofs, xshift=1.2cm] (zf) {${\it z_f}$} ; %
    \node[latent, above=0.5cm of ofs, xshift=-1.2cm] (ws) {$\omega_{s}$} ; %
    \node[const, right=1cm of zf, yshift=0.5cm] (beta1) {$\beta_1$} ; %
    \node[const, right=1cm of zf, yshift=-0.5cm] (beta0) {$\beta_0$} ; %
    \node[const, left=1cm of ws, yshift=0.5cm] (vs) {$\nu_s$} ; %
    \node[const, left=1cm of ws, yshift=-0.5cm] (t2s) {$\tau^2_s$} ; %
    \edge {beta1,beta0} {zf} ; %
    \edge {t2s,vs} {ws} ; %
    \edge {ws,zf} {ofs} ; %
    \plate [inner sep=0.3cm] {f} {(zf)(ofs)} {$F$} ; %
    \plate [inner sep=0.3cm] {} {(ws)(ofs)(f.north west)(f.south west)} {$S$} ; %
  \end{tikzpicture}
  \Description{Fact verification model}
  \caption{Plate diagram of the fact verification model}
  \label{fig:plate}
\end{figure}

The plate diagram of our model is illustrated in Figure~\ref{fig:plate}. For each fact $f$, we model its probability of being true as a latent random variable $z_f$, and generate $z_f$ from a beta distribution: $z_f \sim \text{Beta}(\beta_1,\beta_0)$, with hyperparameter $\bm{\beta}=(\beta_1,\beta_0)$. Beta distribution is a family of continuous probability distributions defined on the interval $[0, 1]$ and often used to describe the distribution of a probability value. $\bm{\beta}$ determines the prior distribution of how likely $f$ is to be true, where $\beta_1$ denotes the prior true count of $f$ and $\beta_0$ denotes the prior false count. The set of all latent truths is denoted by $Z$. Once $z_f$ is calculated, the label $l_f$ of $f$ can be determined by a threshold $\epsilon$, i.e., if $z_f \ge\epsilon$, $l_f = \text{True}$, and False otherwise.

For each source $s$, we model its error variance $\omega_s$ by the scaled inverse chi-squared distribution: $\omega_s \sim \text{Scale-inv-}\chi^2(\nu_s,\tau^2_s)$, with hyperparameters $\nu_s$ and $\tau^2_s$ to encode our belief that source $s$ has labeled $\nu_s$ facts with variance $\tau^2_s$. The set of error variance for all sources is denoted by $W$. We use the scaled inverse chi-squared distribution for two main reasons. First, it can conveniently handle the effect of sample size to tackle the problem brought by the long-tail phenomenon of source claims \cite{CATD}. Second, we use it to keep the scalability of model inference, as it is a conjugate prior for the variance parameter of normal distribution~\cite{Bayesian}.

For each claim $c$ of fact $f$ and source $s$, we assume that the observation $o_{fs}$ is drawn from a normal distribution: $o_{fs} \sim \mathcal{N}(z_f,\omega_s)$, with mean $z_f$ and variance $\omega_s$. The set of observations is denoted by $O$. We believe that the observations are likely to be centered around latent truth and influenced by source quality. Errors, which are the differences between claims and truths, may occur in every source. If a source is unreliable, the observations that it claims would have a wide spectrum and deviate from the latent truths.

\subsection{Truth Inference}

Given the observed claim data, we infer the truths of facts with our probabilistic graphical model. Given hyperparameters $\bm{\beta},\nu_s$ and $\tau^2_s$, the complete likelihood of all observations and latent variables is
\begin{align}
\label{eq:likelihood}
\resizebox{\columnwidth}{!}{$\begin{aligned}
&   \mathrm{Pr}(Z,W,O \,|\, \bm{\beta},\nu_s,\tau^2_s) \\
&= \prod_{f \in F}\mathrm{Pr}(z_f \,|\, \bm{\beta}) \Big(\prod_{s\in S}\prod_{f\in F_s}\mathrm{Pr}(o_{fs} \,|\, z_f,\omega_{s})\Big) \prod_{s\in S}\mathrm{Pr}(\omega_{s} \,|\, \nu_s,\tau^2_s) \\
&= \prod_{f \in F}\text{Beta}{(z_f \,|\, \bm{\beta})} \prod_{s \in S} \Big( \prod_{f \in F_s}\mathcal{N} {(o_{fs} \,|\, z_f,\omega_{s})} \, \text{Scale-inv-}\chi^2(\omega_{s} \,|\, \nu_s, \tau^2_s) \Big),
\end{aligned}$}
\end{align}
where $F, S$ denote the sets of facts and sources, respectively. $F_s$ is the set of facts claimed in source $s$.

We want to find an assignment of latent truths that maximizes the joint probability, i.e., the maximum-a-posterior estimate for $Z$:
\begin{align}
Z^* = \argmax_Z \int \mathrm{Pr}(Z,W,O \,|\, \bm{\beta},\nu_s,\tau^2_s) \, dW.
\end{align}
Note that this derivation holds as $O$ is the observed claim data.

Based on the conjugacy of exponential families, in order to find  $Z^*$, we can directly integrate out $W$ in Eq.~(\ref{eq:likelihood}):
\begin{align}
\label{eq:conjugacy}
\resizebox{\columnwidth}{!}{$\begin{aligned}
& \mathrm{Pr}(Z,O \,|\, \bm{\beta},\nu_s,\tau^2_s) \\
&= \prod_{f \in F} \text{Beta}(z_f \,|\, \bm{\beta}) \prod_{s \in S} \int {\prod_{f \in F_s} \mathcal{N}(o_{fs} \,|\, z_f,\omega_s) \, \text{Scale-inv-}\chi^2(\omega_{s} \,|\, \nu_s,\tau^2_s) } \, d{\omega_s} \\
&\propto \prod_{f \in F} z_f^{\beta_1-1} (1-z_f)^{\beta_0-1} \prod_{s \in S} \Big( \frac{\nu_s\tau^2_s+\sum_{f \in F_s} (z_f-o_{fs})^2}{2} \Big)^{-\frac{\nu_s+|F_s|}{2}}.
\end{aligned}$}
\end{align}

Therefore, the goal becomes to maximize Eq.~(\ref{eq:conjugacy}) w.r.t. $Z$, which is equivalent to minimizing the negative log likelihood:
\begin{align}
\begin{aligned}
& - \log \mathrm{Pr}(Z,O \,|\, \bm{\beta},\nu_s,\tau^2_s) \\
& \propto \sum_{f \in F} \Big( (1-\beta_1) \log z_f + (1-\beta_0) \log(1-z_f) \Big) \\
& \quad + \sum_{s \in S} \frac{\nu_s+|F_s|}{2} \log \frac{\nu_s\tau^2_s+\sum_{f \in F_s} (z_f-o_{fs})^2}{2}.
\end{aligned}
\end{align}

Now, we can apply a gradient descent method to optimize this negative likelihood and infer the unknown latent truths.

\subsection{Hyperparameter Setting and Source Reliability Estimation}
\label{subsect:hyper}

In this section, we describe how to set hyperparameters and how to estimate source reliability, using \emph{prior truths} and observed data.

Intuitively, the error variances of different sources should depend on the quality of observations in the sources, rather than set them as constants regardless of sources. As aforementioned, $o_{fs}$ is drawn from $\mathcal{N}(z_f,\omega_s)$. As the sum of squares of standard normal distributions has the chi-squared distribution~\cite{ChiSquared}, we have
\begin{align}
\frac{\sum_{f \in F_s} (z_f-o_{fs})^2} {{\omega}_s} \sim \chi^2(|F_s|).
\end{align}

Therefore, we have $\omega_{s} \sim \text{Scale-inv-}\chi^2(|F_s|, \hat{\sigma}^2_s)$, where ${\hat{\sigma}^2_s}$  is the sample variance calculated as follows:
\begin{align}
\label{eq:sigma}
\hat{\sigma}^2_s = \frac{1}{|F_s|} {\sum_{f \in F_s} (z_f-o_{fs})^2}.
\end{align}

So, we can set $\nu_s = |F_s|$ and $\tau^2_s = \hat{\sigma}^2_s$ to encode that $s$ has already provided $\nu_s$ observations with average squared deviation $\tau^2_s$. 

Calculating $\hat{\sigma}^2_s$ needs the truths of facts that are claimed in $s$. The most common way is to estimate them by majority voting, which lets $\hat{z}_f = \frac{1}{|S_f|} \sum_{s \in S_f} o_{fs}$, where ${S_f}$ denotes the set of sources that claims ${f}$. A better way is to exploit prior knowledge (i.e., existing facts in the KB) to guide truth inference. Here, we use prior truths derived from a subset of \emph{popular entities} to guide source reliability estimation. For each prior truth in the KB, we directly fix $\hat{z}_f = 1.0$. Besides, we leverage the prior truths to predict whether a property is single-valued or multi-valued by analyzing how popular entities use the property. If the property is single-valued, we only label the fact with the highest probability as correct finally, otherwise the correct facts are determined by threshold $\epsilon$.

Furthermore, we find that $\hat{\sigma}^2_s$ may not accurately reveal the real variance of a source when $|F_s|$ is very small, as many sources have very few claims in the real world, which further causes imprecise truth inference. To solve this issue, we adopt the estimation proposed in~\cite{CATD}, which uses the upper bound of the ${(1-\alpha)}$ confidence interval of sample variance as an estimator, namely,
\begin{align}
\tau^2_s = \frac {\sum_{f \in F_s} (z_f-o_{fs})^2} {\chi^2_{(\alpha/2,|F_s|)}}.
\end{align}

After we obtain the inferred truths of facts using the proposed model, the posterior source quality can be calculated by treating the truths as observed data. The maximum-a-posterior estimate of source reliability has a closed-form solution:
\begin{align}
\text{reliability}(s) \propto \frac {1}{\mathbb{E}[{\omega}_s]} \propto \frac {\nu_s+|F_s|} {\nu_s\tau^2_s+\sum_{f \in F_s} (z_f-o_{fs})^2}.
\end{align}

\begin{figure}
\centering
\includegraphics[width=\columnwidth]{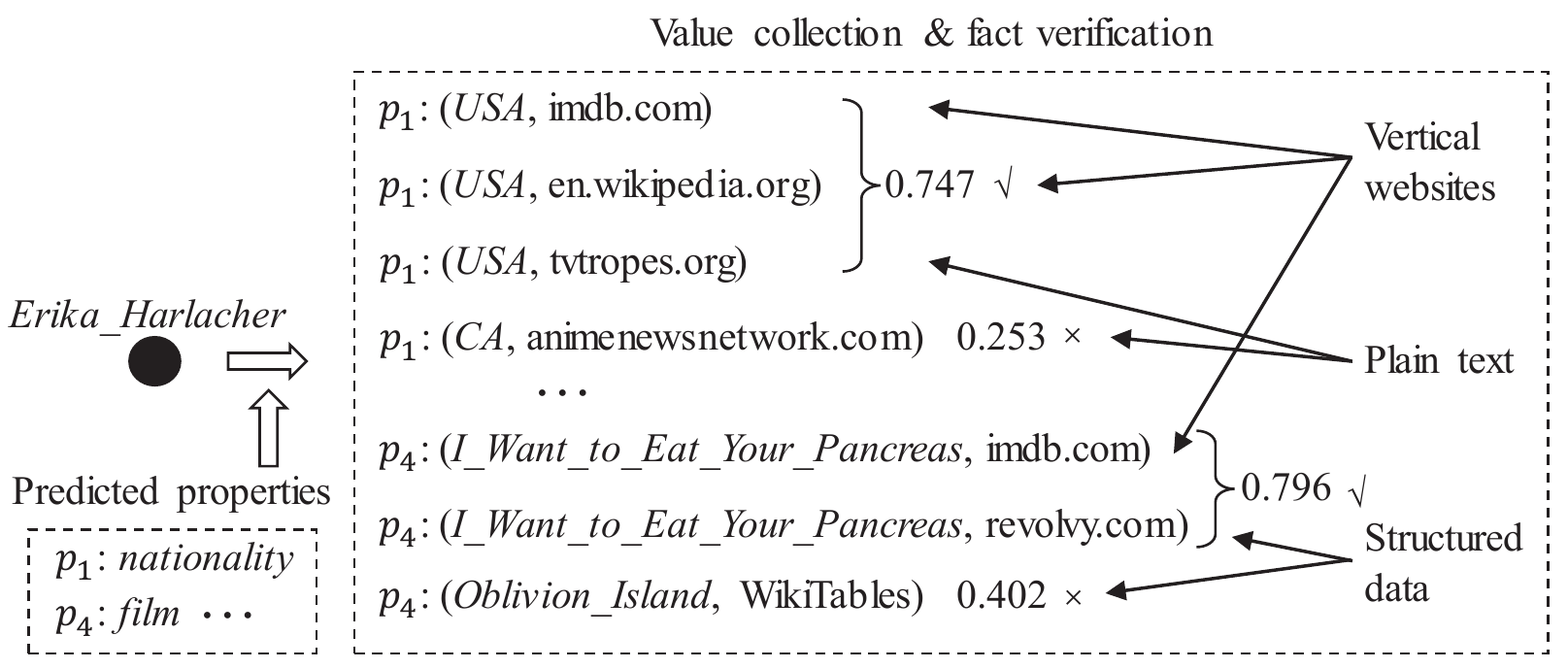}
\Description{Example 2}
\caption{Example of value collection and fact verification}
\label{fig:example2}
\end{figure}

\begin{example}
Recall the example in Figure~\ref{fig:example1}. The attention-based GNN model predicts a few properties like \textit{people.person.nationality} and \textit{film.performance.film}. For each property, many possible values are found from heterogeneous Web sources (see Figure~\ref{fig:example2}). During fact verification, identical facts from different sources are merged, but each source has its own observations. Finally, the nationality of \textit{Erika\_Halacherl} (\textit{USA}, $z_f=0.747$) is correctly identified from others and some films that she dubbed are found as well. $\Box$
\end{example}

Finally, we add the verified facts back to the KB. For a relation fact, we use the same aforementioned entity linking method to link the value to an existing entity in the KB. If the entity is not found, we create a new one with the range of the relation as its type. If this relation has no range, we assign the type from NER to it. For an attribute fact, we simply create a literal for the value.

\section{Experiments and Results}
\label{sect:exp}

\setcounter{footnote}{7}

We implemented our approach, called OKELE, on a server with two Intel Xeon Gold 5122 CPUs, 256 GB memory and a NVIDIA GeForce RTX 2080 Ti graphics card. The datasets, source code and gold standards are accessible online.\footnote{\url{https://github.com/nju-websoft/OKELE/}}

\subsection{Synthetic Experiment}
\label{subsect:synthetic}

\subsubsection{Dataset Preparation} 

Our aim in this experiment was twofold. First, we would like to conduct module-based evaluation of OKELE and compare it with related ones in terms of property prediction, value extraction and fact verification. Second, we wanted to evaluate various (hyper-)parameters in OKELE and use the best in the real-world experiment. As far as we know, there exists no benchmark dataset for open knowledge enrichment yet.

We used Freebase as our KB, because it is famous and widely-used. It is no longer updated, making it more appropriate for our problem, as there is a larger difference against current Web data. We chose 10 classes in terms of popularity (i.e., entity numbers) and familiarity with people. For each class, we created a dataset with 1,000 entities for training, 100 ones for validation and 100 for test, all of which were randomly sampled without replacement from top 20\% of entities by filtered property numbers \cite{EasyFreebase}. Table~\ref{tab:dataset1} lists statistics of the sampled data. 

We leveraged these entities to simulate long-tail entities, based on the local closed world assumption \cite{Facty,Profile} and the leave-$n$-out strategy~\cite{PredSug}. For each entity, we randomly kept five properties (and related facts) and removed the others from the entity. Then, the removed properties and facts were treated as the ground truths in the experiment. We argue that this evaluation may be influenced by the incompleteness of KBs. But, it can be carried out at large scale and without human judgment. Also, since we used popular entities, the problem of incompleteness is not severe.

\begin{table}
\centering
\caption{Statistics of the synthetic dataset}
\label{tab:dataset1}
\begin{tabular}{|l|ccr|}
\hline	\multirow{2}{*}{Classes} & \# Candidate & \# Properties & \# Facts \\
		\cline{3-4} & properties & \multicolumn{2}{c|}{per test entity} \\
\hline	film.actor & 1,241 & 67.5 & 1,167.7 \\
		music.album & \ \, 414 & 22.5 & 92.5 \\
		book.book & \ \, 510 & 22.0 & 66.7 \\
		architecture.building & \ \, 755 & 25.5 & 258.6 \\
		medicine.drug & \ \, 226 & 22.5 & 107.4 \\
		film.film & \ \, 661 & 55.4 & 875.3 \\
		food.food & \ \, 364 & 19.2 & 199.7 \\
		geography.mountain & \ \, 114 & 13.3 & 38.3 \\
		boats.ship & \ \, 131 & 13.0 & 31.7 \\
		computer.software & \ \, 354 & 13.7 & 102.8 \\
\hline
\end{tabular}
\end{table}

\subsubsection{Experiment on Property Prediction} 

Below, we describe the experimental setting and report the results.

\textbf{Comparative models.} We selected three categories of models for comparison: (i) the property mining models designed for KBs, (ii) the traditional models widely-used in recommender systems, and (iii) the deep neural network based recommendation models. For each category, we picked several representatives, which are briefly described as follows. Note that, for all of them, we strictly used the (hyper-)parameters suggested in their papers.
For the first category, we chose the following three models:
\begin{itemize}
\item \emph{Popularity-based}, which ranks properties under a class based on the number of entities using each property. 

\item \emph{Predicate suggester in Wikidata}~\cite{PredSug}, which ranks and suggests candidate properties based on association rules.

\item \emph{Obligatory attributes}~\cite{Obligatory}, which recognizes the obligatory (i.e., not optional) properties of every class in a KB.
\end{itemize}

For the second category, we chose the following three models:
\begin{itemize}
\item \emph{User-KNN and item-KNN}~\cite{KNN}, which are two standard collaborative filtering models. A binary matrix is created based on which entity using which property. 

\item \emph{eALS}~\cite{eALS}, which is a state-of-the-art model for item recommendation, based on matrix factorization.
\end{itemize}

For the last category, we picked three models all in \cite{NeuMF}:
\begin{itemize}
\item \emph{Generalized matrix factorization (GMF)}, which is a full neural treatment of collaborative filtering. It uses a linear kernel to model the latent feature interactions of items and users.

\item \emph{MLP}, which adopts a non-linear kernel to model the latent feature interactions of items and users.

\item \emph{NeuMF}, which is a very recent matrix factorization model with the neural network architecture. It combines GMF and MLP by concatenating their last hidden layers to model the complex interactions between items and users.
\end{itemize}

For OKELE, we searched the initial learning rate in ${\{0.001,0.01,}$ ${0.1\}}$, $\alpha_1$ and $\alpha_2$ in $\{0.1,0.2,0.3,0.4,0.5\}$, the number of entity neighbors $k$ in $\{50,100,200\}$, the number of hidden layers $l$ in $\{1,2,3\}$, the dimension of embeddings and latent vectors $d_1$ in $\{8,16,32\}$, the size of attentions $d_2$ in $\{8,16,32\}$ and the number of negative examples per positive in $\{4,8,16\}$. The final setting is: learning $\text{rate}=0.01$, $\alpha_1=\alpha_2 = 0.3$, $k=100$, $l=1$, both $d_1$ and $d_2=16$, and negative examples per $\text{positive}=4$. Besides, the batch size is 512 and the epoch is 100. The activation function is SeLU(). The model parameters were  initialized with a Gaussian distribution ($\text{mean}=0, \text{SD}=1.0$) and optimized by Adam~\cite{Adam}.

\textbf{Evaluation metrics.} Following the conventions, we employed precision@$m$, normalized discounted cumulative gain (NDCG) and mean average precision (MAP) as the evaluation metrics.

\textbf{Results.} Table~\ref{tab:pp_results} lists the results of the comparative models and OKELE on property prediction. We have three main findings: (i) the models in the second and last categories generally outperform those in the first category, demonstrating the necessity of modeling customized properties for entities, rather than just mining generic properties for classes; (ii) GMF obtains better results than others including the state-of-art model eALS, which shows the power of neural network models in recommendation. Note that MLP underperforms GMF, which is in accord with the conclusion in \cite{NeuMF}. However, we find that NeuMF, which ensembles GMF and MLP, slightly underperforms GMF, due to the non-convex objective function of NeuMF and the relatively poor performance of MLP; and (iii) OKELE is consistently better than all the comparative models. Compared to eALS and GMF, OKELE integrates an advanced GNN architecture to capture complex interactions between entities and properties. Moreover, it uses the attention mechanism during aggregation to model different strengths of the interactions.

\begin{table}
\centering
\caption{Results of property prediction}
\label{tab:pp_results}
\begin{tabular}{|l|cc|cc|c|}
\hline	\multirow{2}{*}{Models} & \multicolumn{2}{c|}{Top-5} & \multicolumn{2}{c|}{Top-10} & \multirow{2}{*}{MAP} \\
		\cline{2-3} \cline{4-5} & Prec. & NDCG & Prec. & NDCG & \\
\hline	Popularity-based & 0.791 & 0.802 & 0.710 & 0.773 & 0.693 \\
		Pred. suggester & 0.879 & 0.892 & 0.783 & 0.857 & 0.758 \\
		Obligatory attrs. & 0.693 & 0.713 & 0.567 & 0.645 & 0.719 \\
\hline	User-KNN & 0.845 & 0.855 & 0.750 & 0.820 & 0.760 \\
          	Item-KNN & 0.881 & 0.890 & 0.800 & 0.875 & 0.804 \\
          	eALS & 0.887 & 0.902 & 0.793 & 0.867 & 0.800 \\
\hline	GMF & \underline{0.906} & \underline{0.916} & \underline{0.805} & \underline{0.880} & \underline{0.814} \\
          	MLP & 0.867 & 0.883 & 0.770 & 0.846 & 0.778 \\
          	NeuMF & 0.901 & 0.915 & 0.797 & 0.875 & 0.800 \\
\hline	OKELE & \textbf{0.926} & \textbf{0.935} & \textbf{0.816} & \textbf{0.891} & \textbf{0.823} \\
\hline	\multicolumn{6}{l}{The best and second best are in \textbf{bold} and \underline{underline}, resp.}
\end{tabular}
\end{table}

\textbf{Ablation study.} We also conducted an ablation study to assess the effectiveness of each module in the property prediction model.  From Table~\ref{tab:ablation1}, we can observe that removing any of these modules would substantially reduce the effect. Particularly, we find that limiting the interactions of entities down to top-$k$ improves the results as controlling the quantity of interactions can filter out noises and concentrate on more informative signals.

\begin{table}
\centering
\caption{Ablation study of the property prediction model}
\label{tab:ablation1}
\begin{tabular}{|l|cc|cc|c|}
\hline	\multirow{2}{*}{Models} & \multicolumn{2}{c|}{Top-5} & \multicolumn{2}{c|}{Top-10} & \multirow{2}{*}{MAP} \\
		\cline{2-3} \cline{4-5} & Prec. & NDCG & Prec. & NDCG & \\
\hline	OKELE & \textbf{0.926} & \textbf{0.935} & \textbf{0.816} & \textbf{0.891} & \textbf{0.823} \\
		w/o attention & 0.909 & 0.921 & 0.805 & 0.880 & 0.808 \\
		w/o top-$k$ ents. & 0.876 & 0.892 & 0.776 & 0.851 & 0.771 \\
		w/o ent. interact. & 0.869 & 0.884 & 0.764 & 0.842 & 0.762 \\
\hline	
\end{tabular}
\end{table}

\subsubsection{Experiment on Value Extraction and Fact Verification} 

In this test, we gave the correct properties to each test entity and compared the facts from the Web with those in the KB (i.e., Freebase). 

\textbf{Comparative models.} We selected the following widely-used models for comparison:
\begin{itemize}
\item \emph{Majority voting}, which regards the fact with the maximum number of occurrences among conflicts as truth.

\item \emph{TruthFinder}~\cite{TruthFinder}, which uses Bayesian analysis to iteratively estimate source reliability and identify truths.

\item \emph{PooledInvestment}~\cite{PooledInvestment}, which uniformly gives a source trustworthiness among its claimed facts, and the confidence of a fact is defined on the sum of reliability from its providers.

\item \emph{Latent truth model (LTM)}~\cite{LTM}, which introduces a graphical model and uses Gibbs sampling to measure source quality and fact truthfulness.

\item \emph{Latent credibility analysis (LCA)}~\cite{LCA}, which builds a strongly-principled, probabilistic model capturing source credibility with clear semantics. 

\item \emph{Confidence-aware truth discovery (CATD)}~\cite{CATD}, which detects truths from conflicting data with long-tail phenomenon. It considers the confidence interval of the estimation.

\item \emph{Multi-truth Bayesian model (MBM)}~\cite{MBM}, which presents an integrated Bayesian approach for multi-truth finding.

\item \emph{Bayesian weighted average (BWA)}~\cite{BWA}, which is a state-of-the-art Bayesian graphical model based on conjugate priors and iterative expectation-maximization inference.
\end{itemize}

Again, for all of them, we strictly followed the parameter setting in their papers. For OKELE, we set $\bm{\beta}=(5,5)$ for all latent truths and chose ${\nu_s},{\tau^2_s}$ following the suggestions in Section~\ref{subsect:hyper}. We used 95\% confidence interval of sample variance, so the corresponding significance level ${\alpha}$ is 0.05. Besides, the threshold $\epsilon$ for determining the labels of facts was set to 0.5 for all the models. 

\textbf{Evaluation metrics.} Following the conventions, we employed precision, recall and F1-score as the evaluation metrics.

\textbf{Results.} Table~\ref{tab:vcfv_results} illustrates the results of value extraction and fact verification. We have four major findings: (i) not surprisingly, OKELE (value extraction) gains the lowest precision but the highest recall as it collects all values from the Web. All the other models conduct fact verification based on these data; (ii) both TruthFinder and LTM achieve lower F1-scores even than majority voting. The reason is that TruthFinder considers the implications between different facts, which would introduce more noises. LTM makes strong assumptions on the prior distributions of latent variables, which fails for the long-tail phenomenon on the Web; (iii) although the precision of MBM is lower than many models, its recall is quite high. The reason is that MBM tends to give high confidence to the unclaimed values, which not only detects more potential truths but also raises more false positives; and (iv) among all the models, OKELE obtains the best precision and F1-score, followed by CATD, since they both handle the challenges of the long-tail phenomenon by adopting effective estimators based on the confidence interval of source reliability. Furthermore, OKELE incorporates prior truths from popular entities to guide the source reliability estimation and truth inference.

\begin{table}
\centering
\caption{Results of value collection and fact verification}
\label{tab:vcfv_results}
\begin{tabular}{|l|ccc|}
\hline	& Prec. & Recall & F1-score \\
\hline	OKELE (value extraction) & 0.222 & \textbf{0.595} & 0.318 \\
\hline	Majority voting & 0.321 & 0.419 & 0.359 \\
		TruthFinder & 0.279 & 0.374 & 0.283 \\
		PooledInvestment & 0.397 & 0.380 & 0.376 \\
		LTM & 0.262 & 0.394 & 0.307 \\
        	LCA & 0.364 & 0.404 & 0.372 \\
		CATD & \underline{0.432} & 0.423 & \underline{0.414} \\
		MBM & 0.340 & \underline{0.539} & 0.401 \\
		BWA & 0.414 & 0.408 & 0.399 \\
\hline	OKELE (fact verification) & \textbf{0.459} & 0.485 & \textbf{0.459} \\
\hline
\end{tabular}
\end{table}

\begin{figure}
\centering
\includegraphics[width=\columnwidth]{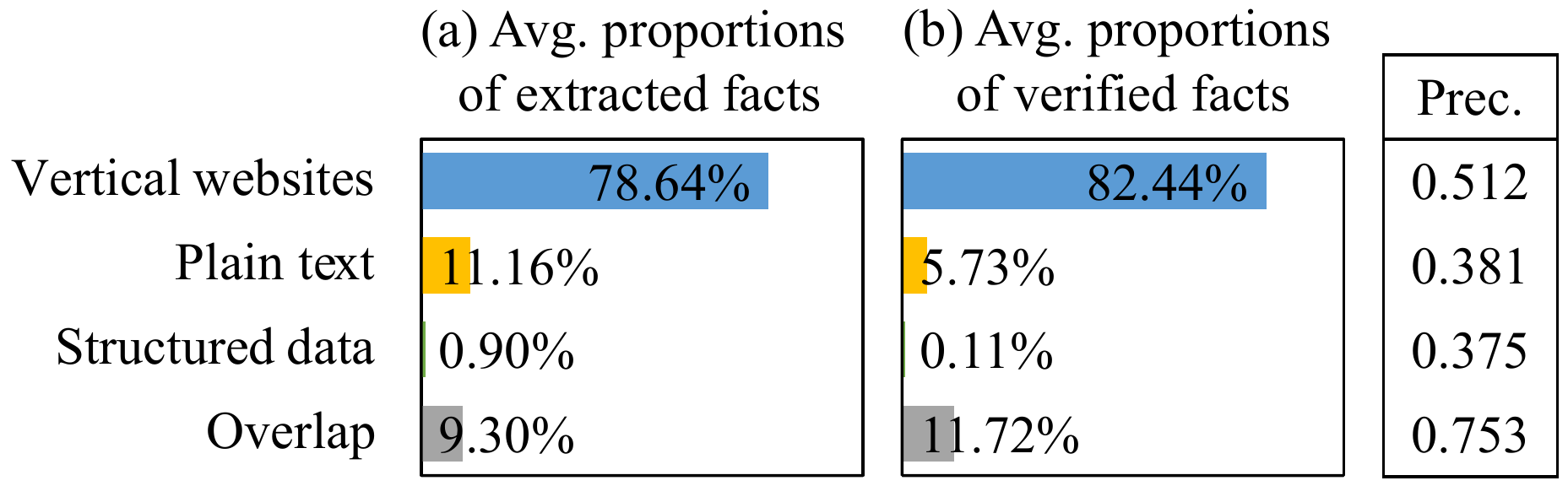}
\Description{Proportions of facts from different source types in the synthetic test}
\caption{Proportions of facts in the synthetic test}
\label{fig:pie1}
\end{figure}

Figure~\ref{fig:pie1} depicts the proportions and precisions of facts from different source types, where ``overlap'' denotes the facts from at least two different source types. We can see from Figure~\ref{fig:pie1}(a) that the number of facts extracted from vertical websites only is the largest. The proportion of facts from structured data only is quite low, as most facts obtained in structured data are also found in other sources. In addition to measuring the proportions of extracted facts, it is important to assess their quality. According to Figure~\ref{fig:pie1}(b), overlap achieves the highest precision. However, most verified facts still come from vertical websites.

\textbf{Ablation study.} We also performed an ablation study to assess the effectiveness of leveraging prior truths from popular entities in the fact verification model. As depicted in Table~\ref{tab:ablation2}, the results show that incorporating prior truths to guide the estimation of source reliability improves the performance.

\begin{table}
\centering
\caption{Ablation study of the fact verification model}
\label{tab:ablation2}
\begin{tabular}{|l|ccc|}
\hline	& Prec. & Recall & F1-score \\
\hline	OKELE (fact verification) & \textbf{0.459} & 0.485 & \textbf{0.459} \\
		w/o prior truths & 0.418 & \textbf{0.499} & 0.438 \\
\hline
\end{tabular}
\end{table}

\subsection{Real-World Experiment}
\label{subsect:real}

\subsubsection{Dataset Preparation and Experimental Setting}

To empirically test the performance of OKELE as a whole, we conducted a real-world experiment on real long-tail entities and obtained the truths of enriched facts by human judgments. For each class, we randomly selected 50 long-tail entities that at least have a name, and the candidate properties are the same as in the synthetic experiment. Table~\ref{tab:dataset2} lists statistics of these samples. For a long-tail entity, OKELE first predicted 10 properties, and then extracted values and verified facts. We hired 30 graduate students in computer software to judge whether a fact is true or false, and a student was paid 25 USD for participation. No one reported that she could not complete the judgments. Each fact was judged by three students to break the tie, and the students should judge it by their own researches like searching the Web. The final ground truths were obtained by voting among the judgments. The level of agreement, measured by Fleiss's kappa~\cite{Kappa}, is 0.812, showing a sufficient agreement. 

As far as we know, there is no existing holistic system that can perform open knowledge enrichment for long-tail entities. So, we evaluated the overall performance of OKELE by comparing with the combination of two second best models GMF\,+\,CATD. As the complete set of facts is unknown, we can only measure precision.

\begin{table}
\centering
\caption{Statistics of the real-world dataset}
\label{tab:dataset2}
\begin{tabular}{|l|cc|l|cc|}
\hline	\multirow{2}{*}{Classes} & \# Props. & \# Facts & \multirow{2}{*}{Classes} & \# Props. & \# Facts \\
		\cline{2-3} \cline{5-6} & \multicolumn{2}{c|}{per test entity} & & \multicolumn{2}{c|}{per test entity} \\
\hline	actor & 4.3 & 10.6 & album & 4.5 & 10.4 \\
		book & 3.5 & 10.6 & building & 3.8 & 13.9 \\
		drug & 3.0 & 14.5 & film & 4.1 & \, 9.3 \\
		food & 2.7 & 14.3 & mountain & 3.7 & 11.5 \\
		ship & 3.5 & \, 8.4 & software & 3.1 & 10.8 \\
\hline
\end{tabular}
\end{table}

\begin{table*}
\centering
\caption{Results of the real-world experiment on different classes}
\label{tab:real_exp}
\begin{tabular}{|l|c|cccccccccc|c|}
\hline	& Models & actor & album & book & building & drug & film & food & mountain & ship & software & Avg. \\
\hline	
\multirow{2}{*}{\# Verified props.} & GMF\,+\,CATD & 280 & 134 & 205 & 218 & 170 & 417 & 65 & 183 & 254 & 207 & 4.27 \\
& OKELE & 264 & 167 & 266 & 209 & 170 & 432 & 70 & 182 & 260 & 199 & 4.44 \\
\hline 
\multirow{2}{*}{\# Verified facts} & GMF\,+\,CATD & 485 & 153 & 228 & 328 & 375 & 722 & 402 & 275 & 303 & 248 & 7.04 \\
& OKELE & 508 & 198 & 418 & 320 & 547 & 1,027 & 615 & 272 & 301 & 247 & 8.91 \\
\hline
\multirow{2}{*}{Precision} & GMF\,+\,CATD & \textbf{0.845} & 0.204 & 0.312 & 0.527 & 0.710 & 0.846 & 0.501 & 0.440 & 0.837 & 0.444 & 0.567 \\
& OKELE & 0.805 & \textbf{0.290} & \textbf{0.464} & \textbf{0.531} & \textbf{0.831} & \textbf{0.890} & \textbf{0.665} & \textbf{0.446} & \textbf{0.870} & \textbf{0.446} & \textbf{0.624} \\
\hline	
\end{tabular}
\end{table*}

\subsubsection{Results}

Table~\ref{tab:real_exp} shows the results of the real-world experiment on different classes. First of all, we see that the results differ among classes. The number of verified facts in class \textit{film} is significantly more than those in other classes and it also holds the highest precision. One reason is that there are several high-quality movie portals containing rich knowledge as people often have great interests in films. In contrast, although people are fond of \textit{albums} as well, OKELE obtains the lowest precision in this class. We find that, since award-winning albums tend to receive more attention, many popular albums in Freebase have award-related properties, which would be further recommended to long-tail albums. However, the majority of long-tail albums have nearly no awards yet. Additionally, albums with very similar names caused disambiguation errors. This also happened in class \textit{food}. OKELE recommended the biological taxonomy properties from natural edible foods to long-tail artificial foods, which have no such taxonomy. In this sense, OKELE may be misguided by using inappropriate popular entities, especially having multiple more specific types. 

The last column of Table~\ref{tab:real_exp} lists the average numbers of verified properties and facts per entity as well as the average precision of 10 classes. Overall, we find that the performance of OKELE is generally good and significantly better than GMF\,+\,CATD. Additionally, the verified facts are from 3,482 Web sources in total. The average run-time per entity is 326.8 seconds, where 24.5\% of the time is spent on network transmission and 40.8\% is spent on NER on plain text. For comparison, we did the same experiment on the synthetic dataset, and OKELE enriched 4.35 properties and 20.59 facts per entity. The average precision, recall and F1-score are 0.479, 0.485 and 0.471, respectively. We owe the precision difference to the incompleteness of the KB. In the synthetic test, we only consider the facts in the KB as correct. Thus, some correct facts from the Web may be misjudged.

We also conducted the module-based evaluation. For property prediction, we measured top-10 precisions w.r.t. properties in the facts that humans judge as correct. The average precisions of OKELE and GMF are 0.497 and 0.428, respectively. For fact verification, we used the same raw facts extracted by OKELE. The average precisions of OKELE and CATD are 0.624 and 0.605, respectively.

\begin{figure}
\centering
\includegraphics[width=\columnwidth]{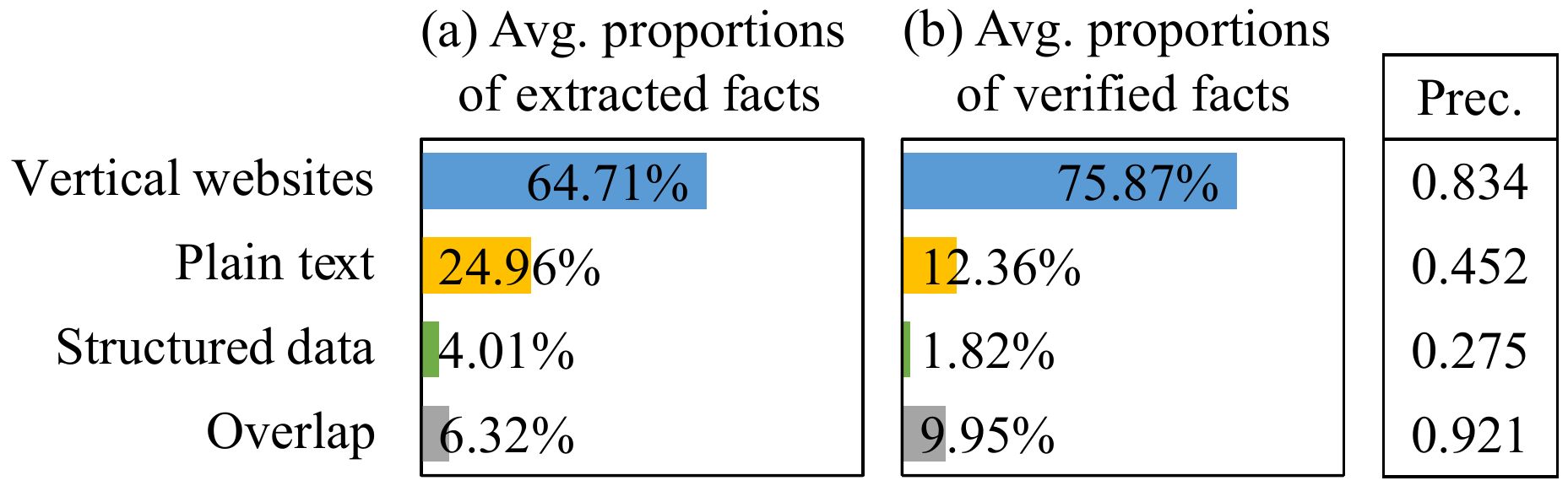}
\Description{Proportions of facts from different source types in the real-world test}
\caption{Proportions of facts in the real-world test}
\label{fig:pie2}
\end{figure}

Figure~\ref{fig:pie2} illustrates the proportions and precisions of facts from different source types in the real-world experiment. Similar to Figure~\ref{fig:pie1}, vertical websites account for the largest proportion and structured data for the least. Overlap still holds the highest precision. However, the proportion of vertical websites declines while the proportion of plain text increases. This is due to that, as compared with popular entities, few people would like to organize long-tail entities into structured or semi-structured knowledge.

\section{Related Work}
\label{sect:work}


\subsection{KB Enrichment}

There is a wide spectrum of works that attempts to enrich KBs from various aspects \cite{KBPSurvey,KGRefineSurvey,KGEmbedSurvey}. According to the data used, we divide existing approaches into \emph{internal} and \emph{external} enrichment.

Internal enrichment approaches focus on completing missing facts in a KB by making use of the KB itself. Specifically, \emph{link prediction} expects one to predict whether a relation holds between two entities in a KB. Recent studies \cite{TransE,ConvE,RotatE} have adopted the embedding techniques to embed entities and relations of a KB into a low-dimensional vector space and predicted missing links by a ranking procedure using the learned vector representations and scoring functions. An exception is ConMask \cite{ConMask}, which learns embeddings of entity names and parts of text descriptions to connect unseen entities to a KB. However, all these studies only target on entity-relations, but cannot handle attribute-values yet. We refer interested readers to the survey \cite{KGEmbedSurvey} for more details. Another line of studies is based upon \emph{rule learning}. Recent notable systems, such as AMIE~\cite{AMIE} and RuleN~\cite{RuleN}, have applied inductive logic programming to mine logical rules and used these rules to deduce missing facts in a KB. In summary, the internal enrichment approaches cannot discover new facts outside a KB and may suffer from the limited information of long-tail entities.

External enrichment approaches aim to increase the coverage of a KB with external resources. The TAC KBP task has promoted progress on information extraction from free text \cite{KBPSurvey}, and many successful systems, e.g., \cite{JointKBP,CSKBP}, use distant supervision together with hand-crafted rules and query expansion \cite{SFOverview}. In addition to text, some KB augmentation works utilize HTML tables \cite{Profile,NovelTable} and embedded markup data \cite{KnowMore} available on the Web. Different from these works that are tailored to specific sources, the goal of this paper is to diversify our Web sources for value extraction. Besides, Knowledge Vault(KV) \cite{KnowledgeVault} is a Web-scale probabilistic KB, in which facts are automatically extracted from text documents, DOM trees, HTML tables and human annotated pages. The main difference between our work and KV is that we aim to enrich long-tail entities in an existing KB, while KV wants to create a new KB and oftentimes regards popular entities.

\subsection{Long-tail Entities}

In the past few years, a few works have begun to pay attention to the long-tail phenomenon in KBs. The work in \cite{DocFLongT} copes with the entity-centric document filtering problem and proposes an entity-independent method to classify vital and non-vital documents particularly for long-tail entities. The work in \cite{LongTNews} analyzes the challenges of linking long-tail entities in news corpora to general KBs. The work in \cite{EmergEntity} recognizes emerging entities in news and other Web streams, where an emerging entity refers to a known long-tail entity in a KB or a new one to be added into the KB. LONLIES \cite{LONLIES} leverages a text corpus to discover entities co-mentioned with a long-tail entity, and generates an estimation of a property-value from the property-value set of co-mentioned entities. However, LONIES needs to give a target property manually and cannot find new property-values that do not exist in the property-value set. The work in \cite{Facty} tackles the problem of knowledge verification for long-tail verticals (i.e., less popular domains). It collects tail-vertical knowledge by crowdsourcing due to the lack of training data, while our work automatically finds and verifies knowledge for long-tail entities from various Web sources. Besides, the work in \cite{LongTWTable} explores the potential of Web tables for extending a KB with new long-tail entities and their descriptions. It develops a different pipeline system with several components, including schema matching, row clustering, entity creation and new detection.

\section{Conclusion}
\label{sect:concl}

In this paper, we introduced OKELE, a full-fledged approach of open knowledge enrichment for long-tail entities. The enrichment process consists of property prediction with attention-based GNN, value extraction from diversified Web sources, and fact verification with probabilistic graphical model, in all of which prior knowledge from popular entities is participated. Our experiments on the synthetic and real-world datasets showed the superiority of OKELE against various competitors. In future, we plan to optimize a few key modules to accelerate the enrichment speed. We also want to study a full neural network-based architecture. 

%
\begin{acks}
This work is supported by the National Natural Science Foundation of China (No. 61872172), the Water Resource Science \& Technology Project of Jiangsu Province (No. 2019046), and the Collaborative Innovation Center of Novel Software Technology \& Industrialization.
\end{acks}

%
\bibliographystyle{ACM-Reference-Format}
\bibliography{main}


\begin{thebibliography}{55}


\ifx \showCODEN    \undefined \def \showCODEN     #1{\unskip}     \fi
\ifx \showDOI      \undefined \def \showDOI       #1{#1}\fi
\ifx \showISBNx    \undefined \def \showISBNx     #1{\unskip}     \fi
\ifx \showISBNxiii \undefined \def \showISBNxiii  #1{\unskip}     \fi
\ifx \showISSN     \undefined \def \showISSN      #1{\unskip}     \fi
\ifx \showLCCN     \undefined \def \showLCCN      #1{\unskip}     \fi
\ifx \shownote     \undefined \def \shownote      #1{#1}          \fi
\ifx \showarticletitle \undefined \def \showarticletitle #1{#1}   \fi
\ifx \showURL      \undefined \def \showURL       {\relax}        \fi
\providecommand\bibfield[2]{#2}
\providecommand\bibinfo[2]{#2}
\providecommand\natexlab[1]{#1}
\providecommand\showeprint[2][]{arXiv:#2}

\bibitem[\protect\citeauthoryear{Bast, B\"{a}urle, Buchhold, and
  Hau{\ss}mann}{Bast et~al\mbox{.}}{2014}]%
        {EasyFreebase}
\bibfield{author}{\bibinfo{person}{Hannah Bast}, \bibinfo{person}{Florian
  B\"{a}urle}, \bibinfo{person}{Bj\"{o}rn Buchhold}, {and}
  \bibinfo{person}{Elmar Hau{\ss}mann}.} \bibinfo{year}{2014}\natexlab{}.
\newblock \showarticletitle{Easy Access to the Freebase Databset}. In
  \bibinfo{booktitle}{\emph{WWW}}. \bibinfo{publisher}{ACM},
  \bibinfo{address}{Seoul, Korea}, \bibinfo{pages}{95--98}.
\newblock


\bibitem[\protect\citeauthoryear{Bhagavatula, Noraset, and Downey}{Bhagavatula
  et~al\mbox{.}}{2013}]%
        {WikiTables}
\bibfield{author}{\bibinfo{person}{Chandra~Sekhar Bhagavatula},
  \bibinfo{person}{Thanapon Noraset}, {and} \bibinfo{person}{Doug Downey}.}
  \bibinfo{year}{2013}\natexlab{}.
\newblock \showarticletitle{Methods for Exploring and Mining Tables on
  Wikipedia}. In \bibinfo{booktitle}{\emph{IDEA}}. \bibinfo{publisher}{ACM},
  \bibinfo{address}{Chicago, IL, USA}, \bibinfo{pages}{18--26}.
\newblock


\bibitem[\protect\citeauthoryear{Bordes, Usunier, Garc\'{i}a{-}Dur\'{a}n,
  Weston, and Yakhnenko}{Bordes et~al\mbox{.}}{2013}]%
        {TransE}
\bibfield{author}{\bibinfo{person}{Antoine Bordes}, \bibinfo{person}{Nicolas
  Usunier}, \bibinfo{person}{Alberto Garc\'{i}a{-}Dur\'{a}n},
  \bibinfo{person}{Jason Weston}, {and} \bibinfo{person}{Oksana Yakhnenko}.}
  \bibinfo{year}{2013}\natexlab{}.
\newblock \showarticletitle{Translating Embeddings for Modeling
  Multi-relational Data}. In \bibinfo{booktitle}{\emph{NIPS}}.
  \bibinfo{publisher}{Curran Associates, Inc.}, \bibinfo{address}{Lake Tahoe,
  NV, USA}, \bibinfo{pages}{2787--2795}.
\newblock


\bibitem[\protect\citeauthoryear{Chen, Feng, Mo, Huang, and Zhao}{Chen
  et~al\mbox{.}}{2014}]%
        {JointKBP}
\bibfield{author}{\bibinfo{person}{Liwei Chen}, \bibinfo{person}{Yansong Feng},
  \bibinfo{person}{Jinghui Mo}, \bibinfo{person}{Songfang Huang}, {and}
  \bibinfo{person}{Dongyan Zhao}.} \bibinfo{year}{2014}\natexlab{}.
\newblock \showarticletitle{Joint Inference for Knowledge Base Population}. In
  \bibinfo{booktitle}{\emph{EMNLP}}. \bibinfo{publisher}{ACL},
  \bibinfo{address}{Doha, Qatar}, \bibinfo{pages}{1912--1923}.
\newblock


\bibitem[\protect\citeauthoryear{Daiber, Jakob, Hokamp, and Mendes}{Daiber
  et~al\mbox{.}}{2013}]%
        {Spotlight}
\bibfield{author}{\bibinfo{person}{Joachim Daiber}, \bibinfo{person}{Max
  Jakob}, \bibinfo{person}{Chris Hokamp}, {and} \bibinfo{person}{Pablo~N.
  Mendes}.} \bibinfo{year}{2013}\natexlab{}.
\newblock \showarticletitle{Improving Efficiency and Accuracy in Multilingual
  Entity Extraction}. In \bibinfo{booktitle}{\emph{I-SEMANTICS}}.
  \bibinfo{publisher}{ACM}, \bibinfo{address}{Graz, Austria},
  \bibinfo{pages}{121--124}.
\newblock


\bibitem[\protect\citeauthoryear{Dettmers, Minervini, Stenetorp, and
  Riedel}{Dettmers et~al\mbox{.}}{2018}]%
        {ConvE}
\bibfield{author}{\bibinfo{person}{Tim Dettmers}, \bibinfo{person}{Pasquale
  Minervini}, \bibinfo{person}{Pontus Stenetorp}, {and}
  \bibinfo{person}{Sebastian Riedel}.} \bibinfo{year}{2018}\natexlab{}.
\newblock \showarticletitle{Convolutional {2D} Knowledge Graph Embeddings}. In
  \bibinfo{booktitle}{\emph{AAAI}}. \bibinfo{publisher}{AAAI Press},
  \bibinfo{address}{New Orleans, LA, USA}, \bibinfo{pages}{1811--1818}.
\newblock


\bibitem[\protect\citeauthoryear{Dong, Gabrilovich, Heitz, Horn, Lao, Murphy,
  Strohmann, Sun, and Zhang}{Dong et~al\mbox{.}}{2014}]%
        {KnowledgeVault}
\bibfield{author}{\bibinfo{person}{Xin~Luna Dong}, \bibinfo{person}{Evgeniy
  Gabrilovich}, \bibinfo{person}{Geremy Heitz}, \bibinfo{person}{Wilko Horn},
  \bibinfo{person}{Ni Lao}, \bibinfo{person}{Kevin Murphy},
  \bibinfo{person}{Thomas Strohmann}, \bibinfo{person}{Shaohua Sun}, {and}
  \bibinfo{person}{Wei Zhang}.} \bibinfo{year}{2014}\natexlab{}.
\newblock \showarticletitle{Knowledge Vault: A Web-Scale Approach to
  Probablistic Knowledge Fusion}. In \bibinfo{booktitle}{\emph{KDD}}.
  \bibinfo{publisher}{ACM}, \bibinfo{address}{New York, NY, USA},
  \bibinfo{pages}{601--610}.
\newblock


\bibitem[\protect\citeauthoryear{Esquivel, Albakour, Martinez, Corney, and
  Moussa}{Esquivel et~al\mbox{.}}{2017}]%
        {LongTNews}
\bibfield{author}{\bibinfo{person}{Jos\'{e} Esquivel}, \bibinfo{person}{Dyaa
  Albakour}, \bibinfo{person}{Miguel Martinez}, \bibinfo{person}{David Corney},
  {and} \bibinfo{person}{Samir Moussa}.} \bibinfo{year}{2017}\natexlab{}.
\newblock \showarticletitle{On the Long-Tail Entities in News}. In
  \bibinfo{booktitle}{\emph{ECIR}}. \bibinfo{publisher}{Springer},
  \bibinfo{address}{Aberdeen, UK}, \bibinfo{pages}{691--697}.
\newblock


\bibitem[\protect\citeauthoryear{Farid, Ilyas, Whang, and Yu}{Farid
  et~al\mbox{.}}{2016}]%
        {LONLIES}
\bibfield{author}{\bibinfo{person}{Mina Farid}, \bibinfo{person}{Ihab~F.
  Ilyas}, \bibinfo{person}{Steven~Euijong Whang}, {and} \bibinfo{person}{Cong
  Yu}.} \bibinfo{year}{2016}\natexlab{}.
\newblock \showarticletitle{LONLIES: Estimating Property Values for Long Tail
  Entities}. In \bibinfo{booktitle}{\emph{SIGIR}}. \bibinfo{publisher}{ACM},
  \bibinfo{address}{Pisa, Italy}, \bibinfo{pages}{1125--1128}.
\newblock


\bibitem[\protect\citeauthoryear{Gal\'{a}rraga, Teflioudi, Hose, and
  Suchanek}{Gal\'{a}rraga et~al\mbox{.}}{2013}]%
        {AMIE}
\bibfield{author}{\bibinfo{person}{Luis~Antonio Gal\'{a}rraga},
  \bibinfo{person}{Christina Teflioudi}, \bibinfo{person}{Katja Hose}, {and}
  \bibinfo{person}{Fabian Suchanek}.} \bibinfo{year}{2013}\natexlab{}.
\newblock \showarticletitle{{AMIE}: Association Rule Mining under Incomplete
  Evidence in Ontological Knowledge Bases}. In \bibinfo{booktitle}{\emph{WWW}}.
  \bibinfo{publisher}{ACM}, \bibinfo{address}{Rio de Janeiro, Brazil},
  \bibinfo{pages}{413--422}.
\newblock


\bibitem[\protect\citeauthoryear{Gelman, Carlin, Stern, Dunson, Vehtari, and
  Rubin}{Gelman et~al\mbox{.}}{2013}]%
        {Bayesian}
\bibfield{author}{\bibinfo{person}{Andrew Gelman}, \bibinfo{person}{John~B.
  Carlin}, \bibinfo{person}{Hal~S. Stern}, \bibinfo{person}{David~B. Dunson},
  \bibinfo{person}{Aki Vehtari}, {and} \bibinfo{person}{Donald~B. Rubin}.}
  \bibinfo{year}{2013}\natexlab{}.
\newblock \bibinfo{booktitle}{\emph{Bayesian Data Analysis}}.
\newblock \bibinfo{publisher}{Chapman and Hall/CRC}, \bibinfo{address}{Boca
  Raton, FL, USA}.
\newblock


\bibitem[\protect\citeauthoryear{Getman, Ellis, Strassel, Song, and
  Tracey}{Getman et~al\mbox{.}}{2018}]%
        {KBPSurvey}
\bibfield{author}{\bibinfo{person}{Jeremy Getman}, \bibinfo{person}{Joe Ellis},
  \bibinfo{person}{Stephanie Strassel}, \bibinfo{person}{Zhiyi Song}, {and}
  \bibinfo{person}{Jennifer Tracey}.} \bibinfo{year}{2018}\natexlab{}.
\newblock \showarticletitle{Laying the Groundwork for Knowledge Base
  Population: Nine Years of Linguistic Resources for {TAC} {KBP}}. In
  \bibinfo{booktitle}{\emph{LREC}}. \bibinfo{publisher}{ELRA},
  \bibinfo{address}{Miyazaki, Japan}, \bibinfo{pages}{1552--1558}.
\newblock


\bibitem[\protect\citeauthoryear{Gunaratna, Thirunarayan, and Sheth}{Gunaratna
  et~al\mbox{.}}{2015}]%
        {FACES}
\bibfield{author}{\bibinfo{person}{Kalpa Gunaratna},
  \bibinfo{person}{Krishnaprasad Thirunarayan}, {and} \bibinfo{person}{Amit
  Sheth}.} \bibinfo{year}{2015}\natexlab{}.
\newblock \showarticletitle{{FACES}: Diversity-Aware Entity Summarization Using
  Incremental Hierarchical Conceptual Clustering}. In
  \bibinfo{booktitle}{\emph{AAAI}}. \bibinfo{publisher}{AAAI Press},
  \bibinfo{address}{Austin, TX, USA}, \bibinfo{pages}{116--122}.
\newblock


\bibitem[\protect\citeauthoryear{Halpin, Herzig, Mika, Blanco, Pound, Thompson,
  and Duc}{Halpin et~al\mbox{.}}{2010}]%
        {Kappa}
\bibfield{author}{\bibinfo{person}{Harry Halpin}, \bibinfo{person}{Daniel~M.
  Herzig}, \bibinfo{person}{Peter Mika}, \bibinfo{person}{Roi Blanco},
  \bibinfo{person}{Jeffrey Pound}, \bibinfo{person}{Henry~S. Thompson}, {and}
  \bibinfo{person}{Thanh~Tran Duc}.} \bibinfo{year}{2010}\natexlab{}.
\newblock \showarticletitle{Evaluating Ad-Hoc Object Retrieval}. In
  \bibinfo{booktitle}{\emph{IWEST}}. \bibinfo{publisher}{CEUR},
  \bibinfo{address}{Shanghai, China}, \bibinfo{pages}{1--12}.
\newblock


\bibitem[\protect\citeauthoryear{He, Liao, Zhang, Nie, Hu, and Chua}{He
  et~al\mbox{.}}{2017}]%
        {NeuMF}
\bibfield{author}{\bibinfo{person}{Xiangnan He}, \bibinfo{person}{Lizi Liao},
  \bibinfo{person}{Hanwang Zhang}, \bibinfo{person}{Liqiang Nie},
  \bibinfo{person}{Xia Hu}, {and} \bibinfo{person}{Tat-Seng Chua}.}
  \bibinfo{year}{2017}\natexlab{}.
\newblock \showarticletitle{Neural Collaborative Filtering}. In
  \bibinfo{booktitle}{\emph{WWW}}. \bibinfo{publisher}{IW3C2},
  \bibinfo{address}{Perth, Australia}, \bibinfo{pages}{173--182}.
\newblock


\bibitem[\protect\citeauthoryear{He, Zhang, Kan, and Chua}{He
  et~al\mbox{.}}{2016}]%
        {eALS}
\bibfield{author}{\bibinfo{person}{Xiangnan He}, \bibinfo{person}{Hanwang
  Zhang}, \bibinfo{person}{Min-Yen Kan}, {and} \bibinfo{person}{Tat-Seng
  Chua}.} \bibinfo{year}{2016}\natexlab{}.
\newblock \showarticletitle{Fast Matrix Factorization for Online Recommendation
  with Implicit Feedback}. In \bibinfo{booktitle}{\emph{SIGIR}}.
  \bibinfo{publisher}{ACM}, \bibinfo{address}{Pisa, Italy},
  \bibinfo{pages}{549--558}.
\newblock


\bibitem[\protect\citeauthoryear{Henaff, Bruna, and LeCun}{Henaff
  et~al\mbox{.}}{2015}]%
        {GNN}
\bibfield{author}{\bibinfo{person}{Mikael Henaff}, \bibinfo{person}{Joan
  Bruna}, {and} \bibinfo{person}{Yann LeCun}.} \bibinfo{year}{2015}\natexlab{}.
\newblock \showarticletitle{Deep Convolutional Networks on Graph-Structured
  Data}.
\newblock \bibinfo{journal}{\emph{CoRR}}  \bibinfo{volume}{abs/1506.05163}
  (\bibinfo{year}{2015}), \bibinfo{pages}{1--10}.
\newblock


\bibitem[\protect\citeauthoryear{Hoffart, Altun, and Weikum}{Hoffart
  et~al\mbox{.}}{2014}]%
        {EmergEntity}
\bibfield{author}{\bibinfo{person}{Johannes Hoffart}, \bibinfo{person}{Yasemin
  Altun}, {and} \bibinfo{person}{Gerhard Weikum}.}
  \bibinfo{year}{2014}\natexlab{}.
\newblock \showarticletitle{Discovering Emerging Entities with Ambiguous
  Names}. In \bibinfo{booktitle}{\emph{WWW}}. \bibinfo{publisher}{ACM},
  \bibinfo{address}{Seoul, Korea}, \bibinfo{pages}{385--396}.
\newblock


\bibitem[\protect\citeauthoryear{Hogg and Craig}{Hogg and Craig}{1978}]%
        {ChiSquared}
\bibfield{author}{\bibinfo{person}{Robert~V. Hogg} {and}
  \bibinfo{person}{Allen~T. Craig}.} \bibinfo{year}{1978}\natexlab{}.
\newblock \bibinfo{booktitle}{\emph{Introduction to Mathematical Statistics}
  (\bibinfo{edition}{4th} ed.)}.
\newblock \bibinfo{publisher}{Macmillan Publishing Co.}, \bibinfo{address}{New
  York, NY, USA}.
\newblock


\bibitem[\protect\citeauthoryear{Kingma and Ba}{Kingma and Ba}{2015}]%
        {Adam}
\bibfield{author}{\bibinfo{person}{Diederik~P. Kingma} {and}
  \bibinfo{person}{Jimmy~Lei Ba}.} \bibinfo{year}{2015}\natexlab{}.
\newblock \showarticletitle{Adam: A method for Stochastic Optimization}. In
  \bibinfo{booktitle}{\emph{ICLR}}. \bibinfo{publisher}{OpenReview.net},
  \bibinfo{address}{San Diego, CA, USA}, \bibinfo{pages}{1--15}.
\newblock


\bibitem[\protect\citeauthoryear{Kipf and Welling}{Kipf and Welling}{2017}]%
        {GCN}
\bibfield{author}{\bibinfo{person}{Thomas~N. Kipf} {and} \bibinfo{person}{Max
  Welling}.} \bibinfo{year}{2017}\natexlab{}.
\newblock \showarticletitle{Semi-supervised Classification with Graph
  Convolutional Networks}. In \bibinfo{booktitle}{\emph{ICLR}}.
  \bibinfo{publisher}{OpenReview.net}, \bibinfo{address}{Toulon, France},
  \bibinfo{pages}{1--14}.
\newblock


\bibitem[\protect\citeauthoryear{Kruit, Boncz, and Urbani}{Kruit
  et~al\mbox{.}}{2019}]%
        {NovelTable}
\bibfield{author}{\bibinfo{person}{Benno Kruit}, \bibinfo{person}{Peter Boncz},
  {and} \bibinfo{person}{Jacopo Urbani}.} \bibinfo{year}{2019}\natexlab{}.
\newblock \showarticletitle{Extracting Novel Facts from Tables for Knowledge
  Graph Completion}. In \bibinfo{booktitle}{\emph{ISWC}}.
  \bibinfo{publisher}{Springer}, \bibinfo{address}{Auckland, New Zealand},
  \bibinfo{pages}{364--381}.
\newblock


\bibitem[\protect\citeauthoryear{Lajus and Suchanek}{Lajus and
  Suchanek}{2018}]%
        {Obligatory}
\bibfield{author}{\bibinfo{person}{Jonathan Lajus} {and}
  \bibinfo{person}{Fabian~M. Suchanek}.} \bibinfo{year}{2018}\natexlab{}.
\newblock \showarticletitle{Are All People Married?: Determining Obligatory
  Attributes in Knowledge Bases}. In \bibinfo{booktitle}{\emph{WWW}}.
  \bibinfo{publisher}{IW3C2}, \bibinfo{address}{Lyon, France},
  \bibinfo{pages}{1115--1124}.
\newblock


\bibitem[\protect\citeauthoryear{Li, Dong, Langen, and Li}{Li
  et~al\mbox{.}}{2017}]%
        {Facty}
\bibfield{author}{\bibinfo{person}{Furong Li}, \bibinfo{person}{Xin~Luna Dong},
  \bibinfo{person}{Anno Langen}, {and} \bibinfo{person}{Yang Li}.}
  \bibinfo{year}{2017}\natexlab{}.
\newblock \showarticletitle{Knowledge Verification for Long-Tail Verticals}.
\newblock \bibinfo{journal}{\emph{Proceedings of the VLDB Endowment}}
  \bibinfo{volume}{10}, \bibinfo{number}{11} (\bibinfo{year}{2017}),
  \bibinfo{pages}{1370--1381}.
\newblock


\bibitem[\protect\citeauthoryear{Li, Li, Gao, Su, Zhao, Demirbas, Fan, and
  Han}{Li et~al\mbox{.}}{2014}]%
        {CATD}
\bibfield{author}{\bibinfo{person}{Qi Li}, \bibinfo{person}{Yaliang Li},
  \bibinfo{person}{Jing Gao}, \bibinfo{person}{Lu Su}, \bibinfo{person}{Bo
  Zhao}, \bibinfo{person}{Murat Demirbas}, \bibinfo{person}{Wei Fan}, {and}
  \bibinfo{person}{Jiawei Han}.} \bibinfo{year}{2014}\natexlab{}.
\newblock \showarticletitle{A Confidence-Aware Approach for Truth Discovery on
  Long-Tail Data}.
\newblock \bibinfo{journal}{\emph{Proceedings of the VLDB Endowment}}
  \bibinfo{volume}{8}, \bibinfo{number}{4} (\bibinfo{year}{2014}),
  \bibinfo{pages}{425--436}.
\newblock


\bibitem[\protect\citeauthoryear{Li, Rubinstein, and Cohn}{Li
  et~al\mbox{.}}{2019}]%
        {BWA}
\bibfield{author}{\bibinfo{person}{Yuan Li}, \bibinfo{person}{Benjamin I.~P.
  Rubinstein}, {and} \bibinfo{person}{Trevor Cohn}.}
  \bibinfo{year}{2019}\natexlab{}.
\newblock \showarticletitle{Truth Inference at Scale: A Bayesian Model for
  Adjudicating Highly Redundant Crowd Annotations}. In
  \bibinfo{booktitle}{\emph{WWW}}. \bibinfo{publisher}{IW3C2},
  \bibinfo{address}{San Francisco, CA, USA}, \bibinfo{pages}{1028--1038}.
\newblock


\bibitem[\protect\citeauthoryear{Lin, Shen, Liu, Luan, and Sun}{Lin
  et~al\mbox{.}}{2016}]%
        {OpenNRE}
\bibfield{author}{\bibinfo{person}{Yankai Lin}, \bibinfo{person}{Shiqi Shen},
  \bibinfo{person}{Zhiyuan Liu}, \bibinfo{person}{Huanbo Luan}, {and}
  \bibinfo{person}{Maosong Sun}.} \bibinfo{year}{2016}\natexlab{}.
\newblock \showarticletitle{Neural Relation Extraction with Selective Attention
  over Instances}. In \bibinfo{booktitle}{\emph{ACL}}.
  \bibinfo{publisher}{ACL}, \bibinfo{address}{Berlin, Germany},
  \bibinfo{pages}{2124--2133}.
\newblock


\bibitem[\protect\citeauthoryear{Lockard, Dong, Einolghozati, and
  Shiralkar}{Lockard et~al\mbox{.}}{2018}]%
        {CERES}
\bibfield{author}{\bibinfo{person}{Colin Lockard}, \bibinfo{person}{Xin~Luna
  Dong}, \bibinfo{person}{Arash Einolghozati}, {and} \bibinfo{person}{Prashant
  Shiralkar}.} \bibinfo{year}{2018}\natexlab{}.
\newblock \showarticletitle{CERES: Distantly Supervised Relation Extraction
  from the Semi-Structured Web}.
\newblock \bibinfo{journal}{\emph{Proceedings of the VLDB Endowment}}
  \bibinfo{volume}{11}, \bibinfo{number}{10} (\bibinfo{year}{2018}),
  \bibinfo{pages}{1084--1096}.
\newblock


\bibitem[\protect\citeauthoryear{Manning, Surdeanu, Bauer, Finkel, Bethard, and
  McClosky}{Manning et~al\mbox{.}}{2014}]%
        {StanfordNLP}
\bibfield{author}{\bibinfo{person}{Christopher Manning}, \bibinfo{person}{Mihai
  Surdeanu}, \bibinfo{person}{John Bauer}, \bibinfo{person}{Jenny Finkel},
  \bibinfo{person}{Steven Bethard}, {and} \bibinfo{person}{David McClosky}.}
  \bibinfo{year}{2014}\natexlab{}.
\newblock \showarticletitle{The {Stanford CoreNLP} Natural Language Processing
  Toolkit}. In \bibinfo{booktitle}{\emph{ACL}}. \bibinfo{publisher}{ACL},
  \bibinfo{address}{Baltimore, MD, USA}, \bibinfo{pages}{55--60}.
\newblock


\bibitem[\protect\citeauthoryear{Mausam}{Mausam}{2016}]%
        {OpenIE}
\bibfield{author}{\bibinfo{person}{Mausam}.} \bibinfo{year}{2016}\natexlab{}.
\newblock \showarticletitle{Open Information Extraction Systems and Downstream
  Applications}. In \bibinfo{booktitle}{\emph{IJCAI}}. \bibinfo{publisher}{AAAI
  Press}, \bibinfo{address}{New York, NY, USA}, \bibinfo{pages}{4074--4077}.
\newblock


\bibitem[\protect\citeauthoryear{Meilicke, Fink, Wang, Ruffinelli, Gemulla, and
  Stuckenschmidt}{Meilicke et~al\mbox{.}}{2018}]%
        {RuleN}
\bibfield{author}{\bibinfo{person}{Christian Meilicke}, \bibinfo{person}{Manuel
  Fink}, \bibinfo{person}{Yanjie Wang}, \bibinfo{person}{Daniel Ruffinelli},
  \bibinfo{person}{Rainer Gemulla}, {and} \bibinfo{person}{Heiner
  Stuckenschmidt}.} \bibinfo{year}{2018}\natexlab{}.
\newblock \showarticletitle{Fine-grained Evaluation of Rule- and
  Embedding-Based Systems for Knowledge Graph Completion}. In
  \bibinfo{booktitle}{\emph{ISWC}}. \bibinfo{publisher}{Springer},
  \bibinfo{address}{Monterey, CA, USA}, \bibinfo{pages}{3--20}.
\newblock


\bibitem[\protect\citeauthoryear{Min, Freedman, and Meltzer}{Min
  et~al\mbox{.}}{2017}]%
        {CSKBP}
\bibfield{author}{\bibinfo{person}{Bonan Min}, \bibinfo{person}{Marjorie
  Freedman}, {and} \bibinfo{person}{Talya Meltzer}.}
  \bibinfo{year}{2017}\natexlab{}.
\newblock \showarticletitle{Probabilistic Inference for Cold Start Knowledge
  Base Population with Prior World Knowledge}. In
  \bibinfo{booktitle}{\emph{EACL}}. \bibinfo{publisher}{ACL},
  \bibinfo{address}{Valencia, Spain}, \bibinfo{pages}{601--612}.
\newblock


\bibitem[\protect\citeauthoryear{Mintz, Bills, Snow, and Jurafsky}{Mintz
  et~al\mbox{.}}{2009}]%
        {DS}
\bibfield{author}{\bibinfo{person}{Mike Mintz}, \bibinfo{person}{Steven Bills},
  \bibinfo{person}{Rion Snow}, {and} \bibinfo{person}{Dan Jurafsky}.}
  \bibinfo{year}{2009}\natexlab{}.
\newblock \showarticletitle{Distant Supervision for Relation Extraction without
  Labeled Data}. In \bibinfo{booktitle}{\emph{ACL}}. \bibinfo{publisher}{ACL},
  \bibinfo{address}{Suntec, Singapore}, \bibinfo{pages}{1003--1011}.
\newblock


\bibitem[\protect\citeauthoryear{Oulabi and Bizer}{Oulabi and Bizer}{2019}]%
        {LongTWTable}
\bibfield{author}{\bibinfo{person}{Yaser Oulabi} {and}
  \bibinfo{person}{Christian Bizer}.} \bibinfo{year}{2019}\natexlab{}.
\newblock \showarticletitle{Extending Cross-Domain Knowledge Bases with Long
  Tail Entities using Web Table Data}. In \bibinfo{booktitle}{\emph{EDBT}}.
  \bibinfo{publisher}{OpenProceedings.org}, \bibinfo{address}{Lisbon,
  Portugal}, \bibinfo{pages}{385--396}.
\newblock


\bibitem[\protect\citeauthoryear{Pasternack and Roth}{Pasternack and
  Roth}{2010}]%
        {PooledInvestment}
\bibfield{author}{\bibinfo{person}{Jeff Pasternack} {and} \bibinfo{person}{Dan
  Roth}.} \bibinfo{year}{2010}\natexlab{}.
\newblock \showarticletitle{Knowing What to Believe (when you already know
  something)}. In \bibinfo{booktitle}{\emph{COLING}}. \bibinfo{publisher}{ACL},
  \bibinfo{address}{Beijing, China}, \bibinfo{pages}{877--885}.
\newblock


\bibitem[\protect\citeauthoryear{Pasternack and Roth}{Pasternack and
  Roth}{2013}]%
        {LCA}
\bibfield{author}{\bibinfo{person}{Jeff Pasternack} {and} \bibinfo{person}{Dan
  Roth}.} \bibinfo{year}{2013}\natexlab{}.
\newblock \showarticletitle{Latent Credibility Analysis}. In
  \bibinfo{booktitle}{\emph{WWW}}. \bibinfo{publisher}{IW3C2},
  \bibinfo{address}{Rio de Janeiro, Brazil}, \bibinfo{pages}{1009--1020}.
\newblock


\bibitem[\protect\citeauthoryear{Paulheim}{Paulheim}{2017}]%
        {KGRefineSurvey}
\bibfield{author}{\bibinfo{person}{Heiko Paulheim}.}
  \bibinfo{year}{2017}\natexlab{}.
\newblock \showarticletitle{Knowledge Graph Refinement: A Survey of Approaches
  and Evaluation Methods}.
\newblock \bibinfo{journal}{\emph{Semantic Web}} \bibinfo{volume}{8},
  \bibinfo{number}{3} (\bibinfo{year}{2017}), \bibinfo{pages}{489--508}.
\newblock


\bibitem[\protect\citeauthoryear{Paulheim}{Paulheim}{2018}]%
        {HowMuchTriple}
\bibfield{author}{\bibinfo{person}{Heiko Paulheim}.}
  \bibinfo{year}{2018}\natexlab{}.
\newblock \showarticletitle{How Much is a Triple? Estimating the Cost of
  Knowledge Graph Creation}. In \bibinfo{booktitle}{\emph{ISWC}}.
  \bibinfo{publisher}{CEUR}, \bibinfo{address}{Monterey, CA, USA},
  \bibinfo{pages}{1--4}.
\newblock


\bibitem[\protect\citeauthoryear{Rao, McNamee, and Dredze}{Rao
  et~al\mbox{.}}{2013}]%
        {EntityLink}
\bibfield{author}{\bibinfo{person}{Delip Rao}, \bibinfo{person}{Paul McNamee},
  {and} \bibinfo{person}{Mark Dredze}.} \bibinfo{year}{2013}\natexlab{}.
\newblock \showarticletitle{Entity Linking: Finding Extracted Entities in a
  Knowledge Base}.
\newblock In \bibinfo{booktitle}{\emph{Multi-source, Multilingual Information
  Extraction and Summarization}}. \bibinfo{publisher}{Springer},
  \bibinfo{address}{Berlin, Heidelberg}, \bibinfo{pages}{93--115}.
\newblock


\bibitem[\protect\citeauthoryear{Razniewski and Weikum}{Razniewski and
  Weikum}{2018}]%
        {KBRecall}
\bibfield{author}{\bibinfo{person}{Simon Razniewski} {and}
  \bibinfo{person}{Gerhard Weikum}.} \bibinfo{year}{2018}\natexlab{}.
\newblock \showarticletitle{Knowledge Base Recall: Detecting and Resolving the
  Unknown Unknowns}.
\newblock \bibinfo{journal}{\emph{ACM SIGWEB Newsletter}}  \bibinfo{volume}{3}
  (\bibinfo{year}{2018}), \bibinfo{pages}{1--9}.
\newblock


\bibitem[\protect\citeauthoryear{Reinanda, Meij, and de~Rijke}{Reinanda
  et~al\mbox{.}}{2016}]%
        {DocFLongT}
\bibfield{author}{\bibinfo{person}{Ridho Reinanda}, \bibinfo{person}{Edgar
  Meij}, {and} \bibinfo{person}{Maarten de Rijke}.}
  \bibinfo{year}{2016}\natexlab{}.
\newblock \showarticletitle{Document Filtering for Long-tail Entities}. In
  \bibinfo{booktitle}{\emph{CIKM}}. \bibinfo{publisher}{ACM},
  \bibinfo{address}{Indianapolis, IN, USA}, \bibinfo{pages}{771--780}.
\newblock


\bibitem[\protect\citeauthoryear{Ritze, Lehmberg, Oulabi, and Bizer}{Ritze
  et~al\mbox{.}}{2016}]%
        {Profile}
\bibfield{author}{\bibinfo{person}{Dominique Ritze}, \bibinfo{person}{Oliver
  Lehmberg}, \bibinfo{person}{Yaser Oulabi}, {and} \bibinfo{person}{Christian
  Bizer}.} \bibinfo{year}{2016}\natexlab{}.
\newblock \showarticletitle{Profiling the Potential of Web Tables for
  Augmenting Cross-domain Knowledge Bases}. In \bibinfo{booktitle}{\emph{WWW}}.
  \bibinfo{publisher}{IW3C2}, \bibinfo{address}{Montr\'{e}al, Canada},
  \bibinfo{pages}{251--261}.
\newblock


\bibitem[\protect\citeauthoryear{Sarwar, Karypis, Konstan, and Riedl}{Sarwar
  et~al\mbox{.}}{2001}]%
        {KNN}
\bibfield{author}{\bibinfo{person}{Badrul Sarwar}, \bibinfo{person}{George
  Karypis}, \bibinfo{person}{Joseph Konstan}, {and} \bibinfo{person}{John
  Riedl}.} \bibinfo{year}{2001}\natexlab{}.
\newblock \showarticletitle{Item-Based Collaborative Filtering Recommendation
  Algorithms}. In \bibinfo{booktitle}{\emph{WWW}}. \bibinfo{publisher}{ACM},
  \bibinfo{address}{Hong Kong, China}, \bibinfo{pages}{285--295}.
\newblock


\bibitem[\protect\citeauthoryear{Shi and Weninger}{Shi and Weninger}{2019}]%
        {ConMask}
\bibfield{author}{\bibinfo{person}{Baoxu Shi} {and} \bibinfo{person}{Tim
  Weninger}.} \bibinfo{year}{2019}\natexlab{}.
\newblock \showarticletitle{Open-World Knowledge Graph Completion}. In
  \bibinfo{booktitle}{\emph{AAAI}}. \bibinfo{publisher}{AAAI Press},
  \bibinfo{address}{New Orleans, LA, USA}, \bibinfo{pages}{1957--1964}.
\newblock


\bibitem[\protect\citeauthoryear{Sun, Deng, Nie, and Tang}{Sun
  et~al\mbox{.}}{2019}]%
        {RotatE}
\bibfield{author}{\bibinfo{person}{Zhiqing Sun}, \bibinfo{person}{Zhi-Hong
  Deng}, \bibinfo{person}{Jian-Yun Nie}, {and} \bibinfo{person}{Jian Tang}.}
  \bibinfo{year}{2019}\natexlab{}.
\newblock \showarticletitle{{RotatE}: Knowledge Graph Embedding by Relational
  Rotation in Complex Space}. In \bibinfo{booktitle}{\emph{ICLR}}.
  \bibinfo{publisher}{OpenReview.net}, \bibinfo{address}{New Orleans, LA, USA},
  \bibinfo{pages}{1--18}.
\newblock


\bibitem[\protect\citeauthoryear{Surdeanu and Ji}{Surdeanu and Ji}{2014}]%
        {SFOverview}
\bibfield{author}{\bibinfo{person}{Mihai Surdeanu} {and} \bibinfo{person}{Heng
  Ji}.} \bibinfo{year}{2014}\natexlab{}.
\newblock \showarticletitle{Overview of the {English} Slot Filling Track at the
  {TAC2014} Knowledge Base Population Evaluation}. In
  \bibinfo{booktitle}{\emph{TAC}}. \bibinfo{publisher}{NIST},
  \bibinfo{address}{Gaithersburg, MD, USA}, \bibinfo{pages}{15}.
\newblock


\bibitem[\protect\citeauthoryear{Tonon, Felder, Difallah, and
  Cudr{\'e}-Mauroux}{Tonon et~al\mbox{.}}{2016}]%
        {Voldermortkg}
\bibfield{author}{\bibinfo{person}{Alberto Tonon}, \bibinfo{person}{Victor
  Felder}, \bibinfo{person}{Djellel~Eddine Difallah}, {and}
  \bibinfo{person}{Philippe Cudr{\'e}-Mauroux}.}
  \bibinfo{year}{2016}\natexlab{}.
\newblock \showarticletitle{Voldemortkg: Mapping schema.org and Web Entities to
  Linked Open Data}. In \bibinfo{booktitle}{\emph{ISWC}}.
  \bibinfo{publisher}{Springer}, \bibinfo{address}{Kobe, Japan},
  \bibinfo{pages}{220--228}.
\newblock


\bibitem[\protect\citeauthoryear{Veli\v{c}kovi\'{c}, Cucurull, Casanova,
  Romero, Li\`{o}, and Bengio}{Veli\v{c}kovi\'{c} et~al\mbox{.}}{2018}]%
        {GAT}
\bibfield{author}{\bibinfo{person}{Petar Veli\v{c}kovi\'{c}},
  \bibinfo{person}{Guillem Cucurull}, \bibinfo{person}{Arantxa Casanova},
  \bibinfo{person}{Adriana Romero}, \bibinfo{person}{Pietro Li\`{o}}, {and}
  \bibinfo{person}{Yoshua Bengio}.} \bibinfo{year}{2018}\natexlab{}.
\newblock \showarticletitle{Graph Attention Networks}. In
  \bibinfo{booktitle}{\emph{ICLR}}. \bibinfo{publisher}{OpenReview.net},
  \bibinfo{address}{Vancouver, Canada}, \bibinfo{pages}{1--12}.
\newblock


\bibitem[\protect\citeauthoryear{Wang, Mao, Wang, and Guo}{Wang
  et~al\mbox{.}}{2017}]%
        {KGEmbedSurvey}
\bibfield{author}{\bibinfo{person}{Quan Wang}, \bibinfo{person}{Zhendong Mao},
  \bibinfo{person}{Bin Wang}, {and} \bibinfo{person}{Li Guo}.}
  \bibinfo{year}{2017}\natexlab{}.
\newblock \showarticletitle{Knowledge Graph Embedding: A Survey of Approaches
  and Applications}.
\newblock \bibinfo{journal}{\emph{IEEE Transactions on Knowledge and Data
  Engineering}} \bibinfo{volume}{29}, \bibinfo{number}{12}
  (\bibinfo{year}{2017}), \bibinfo{pages}{2724--2743}.
\newblock


\bibitem[\protect\citeauthoryear{Wang, Sheng, Fang, Yao, Xu, and Li}{Wang
  et~al\mbox{.}}{2015}]%
        {MBM}
\bibfield{author}{\bibinfo{person}{Xianzhi Wang}, \bibinfo{person}{Quan~Z.
  Sheng}, \bibinfo{person}{Xiu~Susie Fang}, \bibinfo{person}{Lina Yao},
  \bibinfo{person}{Xiaofei Xu}, {and} \bibinfo{person}{Xue Li}.}
  \bibinfo{year}{2015}\natexlab{}.
\newblock \showarticletitle{An Integrated Bayesian Approach for Effective
  Multi-Truth Discovery}. In \bibinfo{booktitle}{\emph{CIKM}}.
  \bibinfo{publisher}{ACM}, \bibinfo{address}{Melbourne, Australia},
  \bibinfo{pages}{493--502}.
\newblock


\bibitem[\protect\citeauthoryear{Yin, Han, and Yu}{Yin et~al\mbox{.}}{2008}]%
        {TruthFinder}
\bibfield{author}{\bibinfo{person}{Xiaoxin Yin}, \bibinfo{person}{Jiawei Han},
  {and} \bibinfo{person}{Philip~S. Yu}.} \bibinfo{year}{2008}\natexlab{}.
\newblock \showarticletitle{Truth Discovery with Multiple Conflicting
  Information Providers on the Web}.
\newblock \bibinfo{journal}{\emph{IEEE Transactions on Knowledge and Data
  Engineering}} \bibinfo{volume}{20}, \bibinfo{number}{6}
  (\bibinfo{year}{2008}), \bibinfo{pages}{796--808}.
\newblock


\bibitem[\protect\citeauthoryear{Yu, Gadiraju, Fetahu, Lehmberg, Ritze, and
  Dietze}{Yu et~al\mbox{.}}{2019}]%
        {KnowMore}
\bibfield{author}{\bibinfo{person}{Ran Yu}, \bibinfo{person}{Ujwal Gadiraju},
  \bibinfo{person}{Besnik Fetahu}, \bibinfo{person}{Oliver Lehmberg},
  \bibinfo{person}{Dominique Ritze}, {and} \bibinfo{person}{Stefan Dietze}.}
  \bibinfo{year}{2019}\natexlab{}.
\newblock \showarticletitle{KnowMore - Knowledge Base Augmentation with
  Structured Web Markup}.
\newblock \bibinfo{journal}{\emph{Semantic Web}} \bibinfo{volume}{10},
  \bibinfo{number}{1} (\bibinfo{year}{2019}), \bibinfo{pages}{159--180}.
\newblock


\bibitem[\protect\citeauthoryear{Zangerle, Gassler, Pichl, Steinhauser, and
  Specht}{Zangerle et~al\mbox{.}}{2016}]%
        {PredSug}
\bibfield{author}{\bibinfo{person}{Eva Zangerle}, \bibinfo{person}{Wolfgang
  Gassler}, \bibinfo{person}{Martin Pichl}, \bibinfo{person}{Stefan
  Steinhauser}, {and} \bibinfo{person}{G\"{u}nther Specht}.}
  \bibinfo{year}{2016}\natexlab{}.
\newblock \showarticletitle{An Empirical Evaluation of Property Recommender
  Systems for Wikidata and Collaborative Knowledge Bases}. In
  \bibinfo{booktitle}{\emph{OpenSym}}. \bibinfo{publisher}{ACM},
  \bibinfo{address}{Berlin, Germany}, \bibinfo{pages}{1--18}.
\newblock


\bibitem[\protect\citeauthoryear{Zhang, Deng, Sun, Wang, Chen, Zhang, and
  Chen}{Zhang et~al\mbox{.}}{2019}]%
        {LongTRelExt}
\bibfield{author}{\bibinfo{person}{Ningyu Zhang}, \bibinfo{person}{Shumin
  Deng}, \bibinfo{person}{Zhanlin Sun}, \bibinfo{person}{Guanying Wang},
  \bibinfo{person}{Xi Chen}, \bibinfo{person}{Wei Zhang}, {and}
  \bibinfo{person}{Huajun Chen}.} \bibinfo{year}{2019}\natexlab{}.
\newblock \showarticletitle{Long-tail Relation Extraction via Knowledge Graph
  Embeddings and Graph Convolution Networks}. In
  \bibinfo{booktitle}{\emph{NAACL-HLT}}. \bibinfo{publisher}{ACL},
  \bibinfo{address}{Minneapolis, MN, USA}, \bibinfo{pages}{3016--3025}.
\newblock


\bibitem[\protect\citeauthoryear{Zhao, Rubinstein, Gemmell, and Han}{Zhao
  et~al\mbox{.}}{2012}]%
        {LTM}
\bibfield{author}{\bibinfo{person}{Bo Zhao}, \bibinfo{person}{Benjamin I.~P.
  Rubinstein}, \bibinfo{person}{Jim Gemmell}, {and} \bibinfo{person}{Jiawei
  Han}.} \bibinfo{year}{2012}\natexlab{}.
\newblock \showarticletitle{A Bayesian Approach to Discovering Truth from
  Conflicting Sources for Data Integration}.
\newblock \bibinfo{journal}{\emph{Proceedings of the VLDB Endowment}}
  \bibinfo{volume}{5}, \bibinfo{number}{6} (\bibinfo{year}{2012}),
  \bibinfo{pages}{550--561}.
\newblock


\end{thebibliography}

\end{document}